\documentclass[12pt]{article}
\pdfoutput=1

\usepackage{graphicx}
\usepackage{latexsym}
\usepackage{epsf}
\usepackage{epsfig}
\usepackage{amssymb,amsmath}
\usepackage[english]{babel}
\usepackage{amsfonts}
\usepackage{amssymb}
\usepackage{graphics}
\usepackage{mathrsfs}
\usepackage{verbatim}
\usepackage{float}
\def\beq{\begin{equation}}
\def\eeq{\end{equation}}
\def\bea{\begin{eqnarray}}
\def\eea{\end{eqnarray}}
\newcommand{\bal}{\begin{align}}
\newcommand{\eal}{\end{align}}
\def\ba{\begin{array}}
\def\ea{\end{array}}
\def\bi{\begin{itemize}}
\def\ei{\end{itemize}}
\def\ben{\begin{enumerate}}
\def\een{\end{enumerate}}
\def\beq{\begin{equation}}
\def\eeq{\end{equation}}
\def\bc{\begin{center}}
\def\ec{\end{center}}
 \def\bt{\begin{table}}
\def\et{\end{table}}
 \def\btb{\begin{tabular}}
\def\etb{\end{tabular}}

\def\co{{\mathcal O}}

\def\mass2{mass${}^2$}

\def\ads{${\mathrm A \mathrm d \mathrm S}_5$}

\def\pa{\partial}

\newcommand{\tr}{\rm Tr}

\newcommand{\cph}{c_\phi}
\newcommand{\sph}{s_\phi}

\newcommand{\ha}{{\hat a}}

\title{
\begin{flushright}
\normalsize{
ANL-HEP-PR-07-34\\
EFI-07-15\\}
\end{flushright}
\vspace*{5mm} \Large\textbf{Gauge-Higgs Unification and Radiative
Electroweak Symmetry Breaking in Warped Extra Dimensions}
\vspace*{1.0cm}
\author{\textbf{Anibal D.~Medina~$^{a,d}$, Nausheen
R.~Shah~$^{b,d}$}\\
\textbf{and Carlos E.M.~Wagner~$^{b,c,d}$}\\
\\[0.5cm]
\normalsize\emph{Department of Astronomy and Astrophysics~$^a$,
Enrico
Fermi Institute~$^b$}\\
\normalsize\emph{and Kavli Institute
for Cosmological Physics~$^c$,}\\
\normalsize\emph{University of Chicago, 5640 S. Ellis Ave.,
Chicago, IL 60637, USA} \\
\normalsize\emph{HEP Division, Argonne National Laboratory, 9700
Cass Ave., Argonne, IL 60439, USA~$^d$}}}

\date{\today}
\begin{document}
\setcounter{page}{0}
\maketitle

\vspace*{0.5cm} \maketitle
\begin{abstract}
We compute the Coleman Weinberg effective potential for the Higgs
field in RS Gauge-Higgs unification scenarios based on a bulk
$SO(5)\times U(1)_X$ gauge symmetry, with gauge and fermion fields
propagating in the bulk and a custodial symmetry protecting the
generation of large corrections to the $T$ parameter and the
coupling of the $Z$ to the bottom quark. We demonstrate that
electroweak symmetry breaking may be realized, with proper
generation of the top and bottom quark masses for the same region of
bulk mass parameters that lead to good agreement with precision
electroweak data in the presence of a light Higgs. We compute the
Higgs mass and demonstrate that for the range of parameters for
which the Higgs boson has Standard Model-like properties, the Higgs
mass is naturally in a range that varies between values close to the
LEP experimental limit and about 160 GeV. This mass range may be
probed at the Tevatron and at the LHC. We analyze the KK spectrum
and briefly discuss the phenomenology of the light resonances
arising in our model.
\end{abstract}

\thispagestyle{empty}

\newpage

\setcounter{page}{1}

\section{Introduction}

High energy physics experiments in recent years have confirmed the
predictions of the Standard Model (SM), a renormalizable, chiral
gauge theory based on the group $SU(3)_c \times SU(2)_L \times
U(1)_Y$. In particular, the low energy dynamics of fermions and
gauge bosons of the theory have been tested with great accuracy. The
origin of masses of these fundamental particles, however, remains a
mystery. In the SM, masses arise through the vacuum expectation
value (vev) of a scalar field doublet, which spontaneously  breaks
the electroweak (EW) symmetry $SU(2)_L \times U(1)_Y  \to
U(1)_{EM}$. A physical, neutral scalar field, the so called Higgs
field, appears in the spectrum. Information on the mass of this
scalar particle may be obtained through the quantum corrections that
it induces to the masses and couplings of the EW gauge bosons.
Consistency of the SM predictions with experimental observations is
improved for small values of the Higgs mass, $m_H$, close to the
current experimental bound, $m_H \geq
114.4$~GeV~\cite{Barate:2003sz}.

There are several aspects of the mechanism of the origin of mass
that demand explanation. On one hand, the scale of the spontaneous
symmetry breaking is governed by the size of the scalar mass
parameter appearing in the Higgs field effective potential, and is
much smaller than the Planck scale, the only known mass scale in
nature, besides the dynamical generated QCD scale $\Lambda_{QCD}$.
Moreover, this scale is associated with a negative value of the
squared Higgs mass. There is, however, no dynamical explanation for
the origin of the Higgs effective potential, or for the associated
breakdown of the EW symmetry. Finally, the hierarchy of the fermion
masses of the different generations remains unexplained.

Warped extra dimensions provide a theoretically attractive framework
for the solution of the hierarchy problem of the SM. For one extra
spatial dimension as in RS1~\cite{Randall:1999ee}, the curvature $k$
in the extra dimension induces a warp factor on the four dimensional
metric, which depends on the position in the extra dimension. The
coordinate point, $x_5 = 0$, is associated with a trivial scale
factor, leading to natural scales of the order of the Planck scale.
The other boundary in the compact dimension, $x_5 = L$, is
associated with natural scales that are smaller by an exponential
factor $\exp(-kL)$. For natural values of $k$ of the order of the
Planck scale and values of $kL \simeq 30$, a Higgs field, located at
$x_5=L$ will naturally acquire a vev of the order of the weak scale.
Neither the origin of the breakdown of the EW symmetry, nor the
hierarchy of fermion masses, however, are explained by these
considerations.

The propagation of fermions and gauge fields in the extra dimension
enables a possible explanation for the fermion mass hierarchy.
Indeed, chirality of the fermion zero modes is ensured by imposing
an orbifold symmetry $S_1/Z_2$. Even fields under this orbifold
symmetry acquire zero modes whose masses arise from the vev of the
Higgs field. For fermion fields, the equations of motion demand that
fields of opposite chirality to those having zero modes are odd
under the orbifold symmetry and therefore have no zero modes.
Finally, localization in the fifth dimension is controlled by a mass
parameter $c$. The variation of $c$ by factors of order one induces
exponential variations in the overlap of the fermion wave functions
with the Higgs field wave function and therefore on the induced
masses for the fermions~\cite{Grossman:1999ra},\cite{Huber:2000ie}.
Third generation left handed chiral quark
fields which couple strongly to the Higgs are associated with a mass
parameter $c < 1/2$, and therefore their zero-modes are localized
towards the so-called infra-red (IR) brane at $x_5 = L$. On the
contrary, first and second generation left handed chiral fermions
have mass parameters $c
> 1/2$ and the zero modes are localized towards the so-called
ultraviolet (UV) brane at $x_5 = 0$.

Gauge-Higgs unification models~\cite{gauge:Higgs:unification}
provide a solution to the last mysterious aspect of the SM, while
leading to a dynamical origin for the Higgs field effective
potential. The gauge symmetry of the SM is extended in the bulk and
broken at the boundaries. The Higgs is associated with the fifth
component of gauge fields in the direction of the broken gauge
symmetry. The gauge bosons associated with the broken symmetry are
odd under the orbifold symmetry and the fifth component is
guaranteed to be even, leading to the presence of massless scalar
fields in the theory, with no potential at tree-level. The
Coleman-Weinberg Higgs potential is then determined by quantum
corrections. For certain values of the fermion mass parameters, this
effective potential leads to the Higgs field acquiring a non-zero
vev and to the proper generation of gauge boson and fermion masses.

Warped extra dimensions have a rich spectrum of four dimensional
Kaluza-Klein (KK) excitations. In particular, after EW symmetry
breaking, the KK modes of the weak gauge fields mix with the
would-be zero modes, inducing important modifications to the masses
and couplings of gauge fields at tree-level. Such large
modifications would render the theory inconsistent for values of the
gauge field's masses at the reach of the
LHC~\cite{Davoudiasl:1999tf}. Recently, it was realized that these
large corrections may be prevented by extending the weak gauge
symmetry in the bulk of the extra dimension to incorporate the
custodial symmetry: $SU(2)_L \times SU(2)_R$~\cite{Agashe:2003zs}.
Moreover, a left-right parity symmetry, $P_{LR}$ is necessary to
prevent large corrections to the bottom-quark
couplings~\cite{Agashe:2006at}. A gauge extension of the EW sector
of the model to $SO(5) \times U(1)_X$ provides a framework in which
gauge-Higgs unification and custodial symmetry are possible. The
Higgs potential is fully calculable at one-loop level and a
prediction for the Higgs boson mass may be obtained.

Consistency of the theory at the quantum level, however, demands
that the possible large corrections to precision electroweak
parameters induced by the fermions of the theory be computed. It was
recently shown that complete consistency may be obtained for certain
values of the mass parameters associated with the fermions of the
third generation~\cite{Carena:2006bn,Carena:2007ua}, provided the
Higgs remains light, as is expected in Gauge-Higgs unification
models.

In this work, we compute the dynamically generated Higgs potential
in models similar to the ones previously analyzed in
Refs.~\cite{Carena:2006bn,Carena:2007ua,Contino:2006qr}. The Higgs
potential is calculated by means of a generalization of the
framework discussed in Ref.~\cite{Falkowski:2006vi} and the
specifics of our model are complementary to the one discussed in
Ref.~\cite{Contino:2006qr}. We find that a consistent breakdown of
the EW symmetry, with a dynamical generation of masses of the gauge
bosons and the third generation quarks may only be obtained for
certain values of the fermion mass bulk parameters. Interestingly
enough, these mass parameters are the same as those demanded by
consistency with experimental observations at the quantum
level~\cite{Agashe:2005dk}. Moreover, the Higgs boson is predicted
to be light, with mass $m_{H}$ in the range to be probed by the
Tevatron and the LHC in the near future.

This article is organized as follows. In section 2 we describe our
5-dimensional (5D) model and derive the gauge boson spectral
functions. Furthermore, we analyze the $W$ gauge boson form factor
at low energy, defining its coupling to the Higgs field. In section
3 we derive the fermion spectral functions, necessary to compute the
loop-induced Higgs effective potential and the fermion mass
spectrum. We also compute the top quark form factor at low energy
and the associated Yukawa coupling. Section 4 deals with the
explicit computation of the effective potential. In section 5 we
present the numerical analysis of the effective potential for the
Higgs and the KK spectrum of the theory. We reserve section 6 for
our conclusions.

\section{5-Dimensional Model and Gauge Fields}
\label{s.5d}

We are interested in a 5D gauge theory with gauge group
$SO(5)\times U(1)_X$. The geometry of our space-time will be that
of RS1~\cite{Randall:1999ee}, with an orbifolded extra spatial
dimension in the interval $x_5 \in [0,L]$. The metric for such a
geometry is given by

\begin{equation} \label{e.wb}
ds^2 = a^2(x_5)\eta_{\mu\nu} dx^\mu dx^\nu - dx_5^2 \, .
\end{equation}
where $a(x_5) = e^{- k x_5}$. The space spanning the fifth dimension
corresponds to a slice of \ads, with branes attached at the two
boundary points: $x_5 = 0$ (UV brane) and $x_5 = L$ (IR brane).

We place our gauge fields, $A_M = A_M^\alpha T^\alpha$ and $B_M$, in
the bulk, where $T^\alpha$ are the hermitian generators of the
fundamental representation of $SO(5)$ and generically
$\tr[T^{\alpha}.T^{\beta}]=C(5)\delta^{\alpha,\beta}$. The explicit
form of the generators~\cite{Agashe:2004rs} are given in Appendix A.

Our fermions $\psi$ also live in the bulk, and they transform under
a representation $t^\alpha$ of $SO(5)$. However, for the moment, we
concentrate only on the gauge content of our model. The
phenomenologically required fermionic content will be discussed in
detail in Section 3.

The 5D action is
\begin{equation}
S_{5D}= \int d^4 x \int_0^Ldx_5 \sqrt{g} \left (-\frac{1}{4g_5^2}\tr
\{ F_{MN}F^{MN} \} -\frac{1}{4 g_X^2} G_{MN}G^{MN} + \bar{\psi} (i
\Gamma^N D_N- M)\psi \right ),\label{5Daction}
\end{equation}
where $D_N = \pa_N - i A_N^\alpha t^\alpha-iB_N$ and $g_5$ and $g_X$
are the 5D dimensionful gauge couplings.

The choice $C(5)=1$ is a convenient choice, since it allows us to
identify the eigenvalues of our generators as the weak isospin, with
the four dimensional coupling given by $g^2=g_5^2/L$. Any other
choice for $C(5)$ may be absorbed into a redefinition of the gauge
fields or the gauge coupling leaving the physics unchanged.

To construct a realistic 4D low energy theory, we will break the
5D $SO(5)\times U(1)_{X}$ gauge symmetry down to the subgroup
$SO(4)\times U(1)_Y=SU(2)_L\times SU(2)_R\times U(1)_Y$ on the IR
brane and to $SU(2)_L\times U(1)_Y$ on the UV brane, where
$Y/2=T^{3_{R}}+Q_{X}$ is the hypercharge and $Q_{X}$ is the
$U(1)_{X}$ associated charge which is accommodated to obtain the
correct hypercharge. We divide the generators of $SO(5)$ as
follows: the generators of $SU(2)_{L,R}$ are denoted by
$T^{a_{L,R}}$ and $t^{a_{L,R}}$, while the generators from the
coset $SO(5)/SO(4)$ are denoted by $T^\ha$ and $t^\ha$.

In order to obtain the correct hypercharge and therefore the right
Weinberg angle $\theta_{W}$, we need to rotate the fields $A^{
3_{R}}_M\in SU(2)_{R}$ and $B_{M}\in
U(1)_{X}$~\cite{Sakamura:2006rf},
\begin{eqnarray}
\begin{array}{c}
\begin{array}{ccccccccccccc}
\begin{pmatrix}
A^{\prime 3_{\rm R}}_M\\
A^Y_M \end{pmatrix} &=& \begin{pmatrix}
\cph & -\sph \\
\sph & \cph
\end{pmatrix}.\begin{pmatrix}
A^{3_{\rm R}}_M\\
B_M \end{pmatrix}
\end{array}
\end{array}
\end{eqnarray}
\begin{equation}
\cph \equiv \frac{g_5}{\sqrt{g_5^2+g_X^2}} ~~,~~
  \sph \equiv \frac{g_X}{\sqrt{g_5^2+g_X^2}} ~~.
\end{equation}
where we will enforce $A^{Y}_\mu$ to have even parity, corresponding
to the hypercharge gauge boson in the 4D low energy limit. From now
on we will drop the prime on $A^{\prime 3_{\rm R}}$, and it will be
understood that $a_{\rm R}$ refers to $1_{\rm R}, 2_{\rm R}$ and
$\prime 3_{\rm R}$.

To implement the breaking of $SO(5)$ on the two branes as stated
above, we impose the following boundary conditions on the gauge
fields:

\begin{eqnarray}
\pa_5 A_\mu^{a_{\rm L}, Y} = A_\mu^{a_{\rm R}, \ha} =
A_5^{a_{\rm L}, Y} &=& 0 \, , \qquad x_5 = 0 \,\label{bc1}\\
\pa_5 A_\mu^{a_{\rm L}, a_{\rm R}, Y} = A_\mu^{\ha}= A_5^{a_{\rm
L}, a_{\rm R}, Y}&=& 0 \, , \qquad x_5 = L \, .\label{bc2}
\end{eqnarray}

These boundary conditions lead to 4D scalars $h^\ha$ originating
from $A_5^\ha$. At tree level, the $h^\ha$'s have a flat potential
and therefore may be thought of as a dual description of Goldstone
bosons arising after spontaneous breaking of the global symmetry
in 4D. Furthermore, $A_\mu^{a_L}$ and $A_\mu^{Y}$ also have zero
modes which will become massive as soon as the $h^\ha$'s develop a
vev, leading ultimately to electroweak symmetry breaking.

Part of our work consists of obtaining the mass spectrum of the
theory in the presence of the vev of $h^\ha$. The Higgs forms a
bidoublet under $SO(4)$, whose doublet component under $SU(2)_{L}$
is given by
$H\propto(h^{\hat{1}}+ih^{\hat{2}},h^{\hat{4}}-ih^{\hat{3}})^{t}$,
with charge 0 under the $U(1)_X$. The $SO(4)$ symmetry in principle
implies that we can choose any one of the components to be the only
non-zero one acquiring a vev. However, we want the scalar Higgs to
be a neutral particle, therefore, we choose $<h^{\hat{4}}>=h$.

We KK-expand the fields~\cite{Falkowski:2006vi},
\begin{eqnarray}
\label{e.kkp}
A_\mu^a(x,x_5) =
\sum_n f_n^a(x_5,h) A_{\mu,n}(x)
&\quad&
A_5^a(x,x_5) = \sum_n \frac{\partial_5 f_n^a(x_5,h)}{ m_n(h)} h_n(x)
\nonumber\\
A_\mu^\ha(x,x_5) =  \sum_n f_n^\ha(x_5,h) A_{\mu,n}(x)
&\quad&
A_5^\ha(x,x_5) =  \frac{C_h}{ a^{2}(x_5)} h^\ha(x)  + \sum_n \frac{\partial_5 f_n^\ha(x_5,h)}{m_n(h)} h_{n}(x)
\end{eqnarray}
where we have put the explicit dependence on $x_{5}$ for the scalars
$h^\ha(x)$. The normalization constant is chosen so that the scalars
are canonically normalized, $C_h = g_5 (\int_0^L a^{-2})^{-1/2}$. We
wish to solve for the KK profiles $f_n(x_5,h)$ so that the 5D action
may be rewritten in terms of a tower of 4D fields $A_n(x)$ whose
kinetic and mass terms, after proper diagonalization, take the form:
\begin{equation}
S_{5D} = \int d^4x \left\{ \frac{1}{2}
(\partial_\mu h^\ha)^2 + \sum_n \left( -\frac{1}{4} [\partial_\mu A_{\nu,n} -
\partial_\nu A_{\mu,n}]^2 + \frac{1}{2} m_n^2(h) A_{\mu,n}^2
\right ) + \dots \right \}
\end{equation}
This is done by solving the equations of motion in the presence of
the scalar vev, and satisfying the boundary conditions implied by
Eqs.~(\ref{bc1}) and (\ref{bc2}).

It turns out that solving the equations of motion in the presence of
$h$ is complicated, as these mix the Neumann and Dirichlet modes.
However, 5D gauge symmetry relates these solutions to solutions with
$h = 0$~\cite{Hosotani:2005nz}. To that end, focusing on the gauge
fields, we perform the following gauge transformation on them
\begin{equation}
\label{e.vrt}
f^\alpha(x_5,h) T^\alpha = \Omega^{-1}(x_5,h) f^\alpha(x_5,0)
T^\alpha \Omega(x_5,h) ,
\end{equation}
where $\Omega(x_5,h)$ is the gauge transformations that removes the
vev of $h$:
\begin{equation}
\label{om.tr}
\Omega(x_5,h) =
\exp\left[-iC_hhT^{4}\int_0^{x_5}dy\,a^{-2}(y)\right] .
\end{equation}

In the $h = 0$ gauge, the gauge KK profiles satisfy the following equation of
motion:
\begin{equation}
\label{geom}
\left (\partial_5^2  + 2 \frac{a'}{a} \partial_5 +  \frac{m_n^2}{a^2} \right )
f_n^\alpha(x_5,0) = 0
\end{equation}
which is the same for both Dirichlet and Neumann modes. We call the
independent solutions $C(x_5,m_n)$ and $S(x_5,m_n)$, which satisfy
the following initial conditions: $C(0,z) = 1$, $C'(0,z) = 0$,
$S(0,z) = 0$ and $S'(0,z) = z$. The explicit form of these solutions
are given by~\cite{Pomarol:1999ad}:
\begin{eqnarray} C(x_5,z) &=& \frac{\pi z}{ 2 k} a^{-1}(x_5) \left [
 Y_0 \left ( \frac{z}{k } \right )      J_1 \left ( \frac{z}{ k a(x_5)} \right )
- J_0 \left (\frac{z}{k} \right )     Y_1 \left (\frac{z}{ k
a(x_5)} \right ) \right ]\\
S(x_5,z) &=&  \frac{\pi z}{ 2 k} a^{-1}(x_5) \left [
 J_1 \left ( \frac{z}{ k } \right )      Y_1 \left ( \frac{z}{ k a(x_5)} \right )
- Y_1 \left ( \frac{z}{ k} \right )     J_1 \left (\frac{z}{ k
a(x_5)} \right ) \right ]
\end{eqnarray}

Note from Eq.~(\ref{om.tr}) that for $x_5 = 0$, $f_n^\alpha(x_5,0)=
f_n^\alpha(x_5,h)$. This implies that the gauge KK profiles
satisfying the UV boundary conditions given in Eq.~(\ref{bc1}), can
be directly written in terms of the basis functions defined above as
\begin{equation}
\begin{array}{cc}
 f_n^{a_{\rm L}}(x_5,0) = C_{n,a_{\rm L}} C(x_5,m_n), \quad & f_n^\ha(x_5,0)
=
C_{n,\ha} S(x_5,m_n)\\
&\\
 f_n^{Y}(x_5,0) = C_{n,Y} C(x_5,m_n), \quad & f_n^{a_{\rm R}}(x_5,0) =
C_{n,a_{\rm R}}S(x_5,m_n) \\
\end{array}
\end{equation}
where the coefficients $C_{n,\alpha}$ are normalization constants.
We can now calculate $f_n^\alpha(x_5,h)$ using Eq.~(\ref{e.vrt}).
The IR boundary conditions give us a system of algebraic equations
for the coefficients $C_{n,\alpha}$. The determinant of this system
of equations must vanish to give us a nontrivial solution, and this
condition in fact gives us the quantization condition for the masses
$m_n(h)$. Using this procedure we find for the gauge bosons:

\begin{eqnarray}
 & &S(L,m_n)S'^3(L,m_n)C'(L,m_n)\times\nonumber\\
 &&\left[C'(L,m_n) S(L,m_n) \left(-2+\sin^2\left(\frac{\lambda_G h}{f_h}\right)\right) - S'(L,m_n)
C(L,m_n)\sin^2\left(\frac{\lambda_G h}{f_h}\right)\right]^{2}\times\nonumber\\
&&\left[C'(L,m_n)
S(L,m_n)\left(-2+(1+\sph^{2})\sin^2\left(\frac{\lambda_G
h}{f_h}\right)\right)-S'(L,m_n)
C(L,m_n)\left(1+\sph^{2}\right)\sin^2\left(\frac{\lambda_G h}{f_h}\right)\right] = 0\nonumber\\
&&
\end{eqnarray}
where the ``Higgs decay constant'' is defined as
\begin{equation}
f_h^2  =  \frac{1}{g_5^2 \int_0^L dy a^{-2}(y) }\label{fh}
\end{equation}
and $\lambda_{G}^2=1/2$. Furthermore, using the Wronskian relation
\begin{equation}
\label{e.wr} S'(x_5,z) C(x_5,z) - C'(x_5,z) S(x_5,z) = z \,
a^{-2}(x_5)
\end{equation}
we can rewrite the quantization condition in a more convenient form:
\begin{eqnarray}
&S(L,m_n)S'^3(L,m_n)C'(L,m_n)\left[2a^{2}_LC'(L,m_n) S(L,m_n)
+m_n\sin^2\left(\frac{\lambda_G h}{f_h}\right)\right]^{2}\times &\nonumber\\
&&\nonumber\\
&\left[2a^{2}_LC'(L,m_n) S(L,m_n)
+(1+\sph^2)m_n\sin^2\left(\frac{\lambda_G h}{f_h}\right)\right]=0&\nonumber\label{gaugedet}\\
&&
\end{eqnarray}

Therefore the KK mass spectrum for the $W$ and $Z$ bosons, with the
correct Weinberg angle given by $\sph^2\simeq
\tan^{2}\theta_{W}\simeq(0.23/0.77)\simeq 0.2987$, is given by the
zeroes of the following equations:

\begin{eqnarray}
\label{gsf} 
1 +  F_{W,Z}(m_n^2) \sin^2\left(\frac{\lambda_G h}{f_h}\right) = 0
\, ,\qquad  \qquad F_{W}(z^2) = \frac{z}{ 2a_L^{2}
C'(L,z) S(L,z)} \nonumber\\
\qquad  \qquad \qquad  \qquad F_{Z}(z^2) = \frac{(1+\sph^2)z}{
2a_L^{2} C'(L,z) S(L,z)} \, .
\end{eqnarray}

We will identify the first zero of these equations with the masses
of $W$ and Z, respectively, and we shall denote the masses of the
first excited states as $m_{W^1}$ and $m_{Z^1}$, associated with the
second zeroes of both equations.

We can gain some physical insight if we look at Eq.~(\ref{gaugedet})
in the limit $h=0$. In that case, this equation reduces to,
\begin{equation}
S^{4}(L,m_n)S'^3(L,m_n)C'^4(L,m_n)=0
\end{equation}
We can identify the zeroes coming from $C'^4(L,m_n)=0$ with the KK
mass spectrum of the 4 gauge bosons belonging to $SU(2)_{L}\times
U(1)_{Y}$ which are even on both the UV and the IR branes. In the
same way, we identify the zeroes of $S'^3(L,m_n)=0$ with the KK
spectrum of the 3 $SU(2)_{R}$ gauge bosons which are odd on the UV
brane and even in the IR brane. $S^{4}(L,m_n)=0$ is identified with
the KK spectrum of the 4 $SO(5)/SO(4)$ gauge bosons which are odd on
both the UV and IR branes. Thus in Eq.~(\ref{gaugedet}) we associate
the photon KK spectrum with the zeroes of $C'(L,m_n)=0$.

\subsection{Gauge Boson Form Factors at Low Energy}

At small momenta, below the scale $\tilde{k}\equiv k\exp(-kL)$, the
form factor for the $W$ gauge bosons can be approximated by
\begin{equation}
\label{e.ffsz} 
F_{W}(-p^2 ) \approx  \frac{g^2 f_h^2}{2p^2};\qquad \qquad g =
\frac{g_5}{\sqrt{L}}  \, .
\end{equation}
From the last equation, we can find an analytic expression for the
W-mass:
\begin{equation}
\label{e.lpm} 
m_W^2  \approx \frac{g^2 f_h^2}{2}\sin^2\left(\frac{\lambda_G
h}{f_h}\right)+\co(m_W^4/\tilde{k}^2) .
\end{equation}
From this expression we can calculate the Higgs-W-W coupling,
$\lambda_{HWW}$, at linear order, by simply taking the derivative of
$m_W^2$ with respect to the vev of $h$:
\begin{equation}
\label{HWW} \lambda_{HWW}=\frac{\partial m_{W}^2 }{\partial
h}=g^{2}\lambda_{G} f_{h}\sin\left(\frac{\lambda_{G}
h}{f_{h}}\right)\cos\left(\frac{\lambda_{G} h}{f_{h}}\right)= g
m_w \cos\left(\frac{\lambda_{G} h}{f_h}\right)\rightarrow gm_{w}
\end{equation}
where we used $\lambda_G=1/\sqrt{2}$ and in the last step note that
in the limit $\lambda_{G} h/f_{h}\ll 1$ we recover the SM Higgs-W-W
coupling. We will discuss the physical implications of the above
limiting case in section 5.

\section{Fermionic KK profiles}

The SM fermions are embedded in full representations of the bulk
gauge group. The presence of the $SU(2)_R$ subgroup of the full bulk
gauge symmetry ensures the custodial protection of the $T$
parameter~\cite{Agashe:2003zs}. In order to have a custodial
protection of the $Zb_L\bar{b}_L$ coupling, the choice
$T^3_R(b_L)=T^3_L(b_L)$ has to be enforced~\cite{Agashe:2006at}. An
economical choice is to let the SM $SU(2)_{L}$ top-bottom doublet
arise from a ${\bf 5}_{2/3}$ of $SO(5)\times U(1)_X$, where the
subscript refers to the $U(1)_X$ charge. As discussed
in~\cite{Carena:2006bn}, putting the SM $SU(2)_{L}$ singlet top in
the same $SO(5)$ multiplet as the doublet, without further mixing,
is disfavored since for the correct value of the top quark mass this
leads to a large negative contribution to the $T$ parameter at one
loop. Hence we let the right-handed top quark arise from a second
${\bf 5}_{2/3}$ of $SO(5)\times U(1)_X$. The right handed bottom can
come from a ${\bf 10}_{2/3}$ that allows us to write the bottom
Yukawa coupling. For simplicity, and because it allows the
generation of the CKM mixing matrix, we make the same choice for the
first two quark generations. We therefore introduce in the quark
sector three $SO(5)$ multiplets per generation as follows:
\begin{eqnarray}
\label{multiplets}
\begin{array}{c}
\begin{array}{ccccccc}
\xi^{i}_{1L} &\sim& Q^{i}_{1L} &=& \begin{pmatrix}
\chi^{u_{i}}_{1L}(-,+)_{5/3} & q^{u_{i}}_L(+,+)_{2/3} \\
\chi^{d_{i}}_{1L}(-,+)_{2/3} & q^{d_{i}}_L(+,+)_{-1/3}
\end{pmatrix} &\oplus& u^{\prime i}_L(-,+)_{2/3}~, \vspace{3mm}
\\
\xi^{i}_{2R} &\sim& Q^{i}_{2R} &=& \begin{pmatrix}
\chi^{u_{i}}_{2R}(-,+)_{5/3} & q^{\prime {u_{i}}}_R(-,+)_{2/3} \\
\chi^{d_{i}}_{2R}(-,+)_{2/3} & q^{\prime {d_{i}}}_R(-,+)_{-1/3}
\end{pmatrix} &\oplus& u^{i}_R(+,+)_{2/3}~,
\end{array}
\end{array}
\vspace{3mm}
\end{eqnarray}
\begin{eqnarray}
\begin{array}{l}
\xi^{i}_{3R} \sim \\
\\
\begin{array}{ccccccccccc}
T^{i}_{1R} = \begin{pmatrix}
\psi^{\prime i}_R(-,+)_{5/3} \\
U^{\prime i}_R(-,+)_{2/3} \\
D^{\prime i}_R(-,+)_{-1/3} \end{pmatrix} \oplus T^{i}_{2R} =
\begin{pmatrix}
\psi^{\prime\prime i}_R(-,+)_{5/3} \\
U^{\prime\prime i}_R(-,+)_{2/3} \\
D^{i}_R(+,+)_{-1/3} \end{pmatrix} \oplus Q^{i}_{3R} =
\begin{pmatrix}
\chi^{u_{i}}_{3R}(-,+)_{5/3} & q^{\prime \prime u_{i}}_R(-,+)_{2/3} \\
\chi^{d_{i}}_{3R}(-,+)_{2/3} & q^{\prime \prime
d_{i}}_R(-,+)_{-1/3}
\end{pmatrix},
\end{array}
\end{array}\nonumber
\end{eqnarray}
where we show the decomposition under $SU(2)_L\times SU(2)_R$, and
explicitly write the $U(1)_{EM}$ charges. The $Q^{i}$s are
bidoublets of $SU(2)_L\times SU(2)_R$, with $SU(2)_L$ acting
vertically and $SU(2)_R$ acting horizontally. The $T^{i}_{1}$'s and
$T^{i}_{2}$'s transform as $({\bf 3}, {\bf 1})$ and $({\bf 1}, {\bf
3})$ under $SU(2)_L\times SU(2)_R$, respectively, while $u^{i}$ and
$u^{\prime i}$ are $SU(2)_L\times SU(2)_R$ singlets. The
superscripts, $i=1,2,3$, label the three generations.

We also show the parities on the indicated 4D chirality, where $-$
and $+$ stands for odd and even parity conditions and the first
and second entries in the bracket correspond to the parities in
the UV and IR branes respectively. Let us stress that while odd
parity is equivalent to a Dirichlet boundary condition, the even
parity is a linear combination of Neumann and Dirichlet boundary
conditions, that is determined via the fermion bulk equations of
motion as discussed below.

 The boundary conditions for the
opposite chirality fermion multiplet can be read off the ones
above by a flip in both chirality and boundary condition,
$(-,+)_L\rightarrow (+,-)_R$ for example. In the absence of mixing
among multiplets satisfying different boundary conditions, the SM
fermions arise as the zero-modes of the fields obeying $(+,+)$
boundary conditions. The remaining boundary conditions are chosen
so that $SU(2)_{L} \times SU(2)_{R}$ is preserved on the IR brane
and so that mass mixing terms, necessary to obtain the SM fermion
masses after EW symmetry breaking, can be written on the IR brane.
Consistency of the above parity assignments with the original
orbifold $Z_2$ symmetry at the IR brane will be discussed in
Appendix B.

As remarked above, the zero-mode fermions can acquire EW symmetry
breaking masses through mixing effects. The most general $SU(2)_L
\times SU(2)_R\times U(1)_X$ invariant mass Lagrangian at the IR
brane -- compatible with the boundary conditions -- is, in the quark
sector,
\begin{equation}
\mathcal{L}_m = 2\delta(x_{5}-L) \Big[ \bar{u}^\prime_L M_{B_1} u_R
+ \bar{Q}_{1L} M_{B_2} Q_{3R} + \bar{Q}_{1L} M_{B_3} Q_{2R} +
\mathrm{h.c.} \Big] ~, \label{localizedmasses}
\end{equation}
where $M_{B_1}$, $M_{B_2}$ and $M_{B_3}$ are dimensionless $3
\times 3$ matrices, and matrix notation is employed. In Appendix B
we will show how these mass terms may be generated from an $SO(5)$
invariant Lagrangian using the spurion formalism. To prevent
negative corrections to the $T$
parameter~\cite{Peskin:1991sw},~\cite{Carena:2006bn,Carena:2007ua},
we will take $M_{B_3}$ to be the null matrix. In Appendix C
however we briefly discuss the modifications of the effective
Higgs potential induced by a non-zero value of $M_{B_{3}}$.
Moreover since the Higgs effective potential is most sensitive to
the third generation, we will concentrate on this one, discarding
family indices.

With the introduction of these brane mixing terms, the different
multiplets are now related via the equations of motion. In the case
of a mass term involving $\bar{\Psi}^1_{\rm L}\Psi^2_{\rm R}$, with
the fields having profile functions $g_{\rm L}$ and $h_{\rm R}$, the
odd parity profile $g_R$ on the IR brane is related to $h_R$:

\begin{equation}
\label{delta} \int^{L+\epsilon}_{L-\epsilon}(\partial_5
g_{R})dx_5=\int^{L+\epsilon}_{L-\epsilon}(2Mh_{R}-g_R)\delta(x_5-L)dx_5.
\end{equation}
The odd parity we demand at the IR brane implies a Dirichlet
boundary condition for the function $g_{R}$. Therefore, we can
rewrite:
\begin{equation}
\label{limit} \lim_{\epsilon\rightarrow 0}
g_{R}(L-\epsilon)=-Mh_{R}(L),
\end{equation}
and, similarly:
\begin{equation}
\label{limit2} \lim_{\epsilon\rightarrow
0}h_{L}(L-\epsilon)=Mg_{L}(L).
\end{equation}
Eq.~(\ref{limit}) and (\ref{limit2}) can now be reinterpreted as the
new boundary conditions for the profiles at the IR brane.

The fermions, like the gauge bosons, can be expanded in their KK
basis:
\begin{equation}
\psi_{L}(x,x_5) = 
\sum_n f_{L,n}(x_5,h) \psi_{L,n}(x), \quad \quad \psi_{R}(x,x_5) =
\sum_n f_{R,n}(x_5,h) \psi_{R,n}(x).
\end{equation}
In the $h=0$ gauge, we redefine $\hat{\psi}=a^{2}(x_{5})\psi$ and we
write our vector-fermionic fields in terms of chiral fields. We can
KK decompose the fermionic chiral components as,
\begin{equation}
\hat{\psi}_{L,R}=\sum_{n=0}^{\infty}\psi^{n}_{L,R}(x^{\mu})\hat{f}_{L,R,n}(x_{5})
\end{equation}
where $\hat{f}$ is normalized by:
\begin{equation} \label{ferm.norm} \int^L_0 dx_5
a^{-1}(x_5)\hat{f}_n\hat{f}_m=\delta_{m,n}.
\end{equation}
Therefore the profile function for the zero mode fermion corresponds
to $a^{-1/2}(x_5)\hat{f}_0$.

From the 5D action, Eq.~(\ref{5Daction}), concentrating on the free
fermionic fields, we can derive the following first order coupled
equations of motion for $\hat{f}_{L,R,n}$,
\begin{equation}
\label{1order}
(\partial_{5}+M)\hat{f}_{R,n}=(z/a(x_{5}))\hat{f}_{L,n}; \quad
(\partial_{5}-M)\hat{f}_{L,n}=-(z/a(x_{5}))\hat{f}_{R,n}
\end{equation}

We see from Eq.~(\ref{1order}) that we can redefine
$\tilde{f}_{R,L,n}=e^{- Mx_{5}}\hat{f}_{R,L,n}$ and relate the
opposite chiral component of the same vector-like field by
$\tilde{f}_{R,n}=(-a(x_{5})/z)\partial_{5}\tilde{f}_{L,n}$. For the
left handed field having Dirichlet boundary conditions on the UV
brane, we can derive a second order equation for the chiral
component $\tilde{f}_{L,n}$:
\begin{equation}
\label{feom} \left [\partial_5^2  + \left (\frac{a'}{ a} + 2M
\right)
\partial_5   + \frac{z^2}{ a^2} \right ] \tilde{f}_{L,n} = 0
\end{equation}
the solution of which we shall call $\tilde{S}_{M}(x_5,z)$, with
boundary conditions $\tilde{S}_{M}(0,z) = 0$, $\tilde{S}_{\pm
M}'(0,z) = z$.

Similarly, we can redefine $\tilde{f}_{R,L,n}=e^{
Mx_{5}}\hat{f}_{R,L,n}$ and then relate the opposite chirality via
$\tilde{f}_{L,n}=(a(x_{5})/z)\partial_{5}\tilde{f}_{R,n}$. If the
right-handed field fulfills Dirichlet boundary conditions on the
UV-brane, we can
write the equation of motion for $\tilde{f}_{R,n}$:
\begin{equation}
\label{feom1} \left [\partial_5^2  + \left (\frac{a'}{ a} - 2M
\right)
\partial_5   + \frac{z^2}{ a^2} \right ] \tilde{f}_{R,n} = 0.
\end{equation}
We shall correspondingly denote the solution to this equation with
$\tilde{S}_{-M}(x_5,z)$, fulfilling the boundary conditions
$\tilde{S}_{-M}(0,z) = 0$, $\tilde{S}_{- M}'(0,z) = z$.

As emphasized before, the solutions of the opposite chirality, are
related to these solutions via the the first order equations of
motion, Eqs.~(\ref{1order}), which we rewrite as
$\dot{\tilde{S}}_{\pm M}(x_{5},z)=\mp(a(x_{5})/z)\partial_5
\tilde{S}_{\pm M}(x_{5},z)$. Here $M=-ck$, and the $\pm$ is chosen
depending on whether we want an odd (-) boundary condition on the UV
brane for the left-handed or right-handed fermionic field. The
solution to Eq.~(\ref{feom}) is given by~\cite{Pomarol:1999ad}:

\begin{equation}
\label{SM}
\tilde{S}_M(x_5,z) = \frac{\pi z}{ 2 k} a^{-c-\frac{1}{
2} }(x_5) \left [
 J_{\frac{1}{2}+c} \left (\frac{z}{ k } \right )      Y_{\frac{1}{ 2}+c} \left ( \frac{z}{ k a(x_5)} \right )
- Y_{\frac{1}{2}+c } \left ( \frac{z}{ k } \right )
J_{\frac{1}{2}+c}  \left ( \frac{z}{ k a(x_5)} \right ) \right ].
\end{equation}
The solution for Eq.~(\ref{feom1}) is given by Eq.~(\ref{SM}), with
the replacement $c\rightarrow -c$.

Before proceeding with the gauge transformations as was done with
the gauge fields, we would like to discuss what is required of the
gauge transformation for the fermions. The boundary conditions are
explicitly given on the physical fermion which is an isospin
eigenvector. Hence, the first step is to identify the different
components of the vector $\psi$ with the doublets, triplets and
singlets as required from Eq.~(\ref{multiplets}). Second, we need to
assign functions to the different components of the multiplets
consistent with the boundary conditions for the fermions given in
Eq.~(\ref{multiplets}). The basis $f_{L,R,n}$ and
$\tilde{f}_{L,R,n}$ are simply related by exponentials, therefore,
the functions $f_{L,R,n}$, given by $S_{\pm M}$ and $\dot{S}_{\pm
M}$, are related to $\tilde{f}_{L,R,n}$ via $S_{\pm M} (\dot{S}_{\pm
M})=a^{-2}(x_5)\exp[\pm M x_{5}] \tilde{S}_{\pm M}
(\dot{\tilde{S}}_{\pm M})$. Odd and even solutions ($S$ or
$\dot{S}$) can then be assigned to the component functions of the
multiplets $\xi_{1L}$, $\xi_{2R}$ and $\xi_{3R}$, depending on the
boundary parities. For example, for a left handed component having
boundary conditions $(-,+)_{L}$ or $(-,-)_L$, since the function is
odd at the UV brane, we assign a solution $S_{M}$ to that component
of the left-handed multiplet. If the boundary condition is
$(+,+)_{L}$, or $(+,-)_{L}$ we need to look at the chiral companion
$(-,-)_{R}$ or $(-,+)_{R}$.
In this case, our right handed fermion is odd on the UV brane. The
relationship between the left and the right handed components are as
defined above via derivatives. Therefore, we assign $\dot{S}_{-M}$
to the left-handed fermion. Following this procedure, we can assign
either $S_{\pm M}$ or $\dot{S}_{\pm M}$ to each component of the
multiplets. This procedure defines our three fermion multiplets in
the $h=0$ gauge as the vector functions:
\begin{eqnarray}
\label{f.exp.bc}
\begin{array}{c}
f_{1,L}(x_{5},0)=\left[\begin{array}{c} C_{1}S_{M_{1}}\\
C_{2}S_{M_{1}}\\ C_{3}\dot{S}_{-M_{1}}\\
C_{4}\dot{S}_{-M_{1}}\\ C_{5}S_{M_{1}}\end{array}\right]\\
\\
\\
f_{2,R}(x_{5},0)=\left[\begin{array}{c} C_{6}S_{-M_{2}}\\
C_{7}S_{-M_{2}}\\ C_{8}S_{-M_{2}}\\
C_{9}S_{-M_{2}}\\ C_{10}\dot{S}_{M_{2}}\end{array}\right]
\end{array} & &
f_{3,R}(x_{5},0)=\left[\begin{array}{c} C_{11}S_{-M_{3}}\\
C_{12}S_{-M_{3}}\\ C_{13}S_{-M_{3}}\\ C_{14}S_{-M_{3}}\\
C_{15}S_{-M_{3}}\\ C_{16}S_{-M_{3}}\\
C_{17}S_{-M_{3}}\\C_{18}S_{-M_{3}}\\ C_{19}S_{-M_{3}}\\
C_{20}\dot{S}_{M_{3}}\end{array}\right]
\end{eqnarray}
where, as for the gauge bosons, the $C_i$ are normalization
constants. The opposite chiralities have opposite boundary
conditions and can be read from these ones.

The boundary conditions as written above, for example, for the
vector components of $f_{1,L}(x_{5},0)$ are the boundary conditions
on the isospin eigenvector with charges $(+1/2,+1/2)$,
$(-1/2,+1/2)$, $(+1/2,-1/2)$, $(-1/2,-1/2)$, $(0,0)$ under
$SU(2)_L\times SU(2)_R$, respectively. However, in the basis of our
generators, the eigenvector with spin $+1/2$ under $SU(2)_{\rm L}$,
for example, is given by:
\begin{equation}
\label{f.exp}
\hat{e}=\frac{1}{\sqrt{2}}\left[\begin{array}{c}i\\-1\\0\\0\\0\end{array}\right].
\end{equation}

To simplify matters, so that the boundary conditions can be taken to
be on the eigenvector:
\begin{equation}
\label{f.can}
\hat{e}=\left[\begin{array}{c}1\\0\\0\\0\\0\end{array}\right]
\end{equation}
the transformation $\Omega$ needs to be in the basis consistent with
Eq.~(\ref{f.can}), or we can also think about this as changing the
fermion vector functions given in Eq.~(\ref{f.exp.bc}) to the basis
of our generators. This means that we have to first understand how
the eigenvectors of $T^{3\rm R,L}$ decompose in the canonical basis
for both the ${\bf 5}_{2/3}$ and the ${\bf 10}_{2/3}$.

%
%
%
%

%
%
For the ${\bf 5}_{2/3}$ fermions, the basis transformation for
$\xi_{1L}$, $\xi_{2R}$ is given by
\begin{eqnarray}
\label{A.mat}
\begin{array}{ccccc}
 A &=& \frac{1}{\sqrt{2}}\begin{pmatrix}
-i & -1 & 0 & 0 & 0 \\
0 & 0 & -i & 1 & 0 \\
0 & 0 & i & 1 & 0 \\
-i & 1 & 0 & 0 & 0 \\
0 & 0 & 0 & 0 & \sqrt{2}
\end{pmatrix}
\end{array}
\end{eqnarray}

For ${\bf 10}_{2/3}$, the transformation matrix for $\xi_{3R}$
is given by
\begin{eqnarray}
\label{B.mat}
\begin{array}{cccccccccc}
B &=& \frac{1}{\sqrt{2}}\begin{pmatrix}
i & 1 & 0 & 0 & 0 & 0 & 0 & 0 & 0 & 0 \\
0 & 0 & -i & 1 & 0 & 0 & 0 & 0 & 0 & 0 \\
0 & 0 & -i & -1 & 0 & 0 & 0 & 0 & 0 & 0 \\
-i & 1 & 0 & 0 & 0 & 0 & 0 & 0 & 0 & 0 \\
0 & 0 & 0 & 0 & -1 & i & 0 & 0 & 0 & 0 \\
0 & 0 & 0 & 0 & 0 & 0 & \sqrt{2} & 0 & 0 & 0 \\
0 & 0 & 0 & 0 & 1 & i & 0 & 0 & 0 & 0 \\
0 & 0 & 0 & 0 & 0 & 0 & 0 & -1 & i & 0 \\
0 & 0 & 0 & 0 & 0 & 0 & 0 & 0 & 0 & \sqrt{2} \\
0 & 0 & 0 & 0 & 0 & 0 & 0 & 1 & i & 0
\end{pmatrix}
\end{array}
\end{eqnarray}

Therefore, once the UV boundary conditions define the vectors as in
Eq.~(\ref{f.exp.bc}), we change bases to be able to implement the
gauge transformation and then transform back so that the IR boundary
conditions can be imposed. Using the transformations defined in
Eq.~(\ref{A.mat}) and (\ref{B.mat}), we can write the final fermion
vector functions in the presence of the vev of $h$ as:
\begin{eqnarray}
& & f_{1,L}(x_{5},h) = A\Omega A^{-1} f_{1,L}(x_{5},0)\\
& & f_{2,R}(x_{5},h) = A\Omega A^{-1} f_{2,R}(x_{5},0)\\
& & f_{3,R}(x_{5},h) = B\Omega B^{-1} f_{3,R}(x_{5},0)
\end{eqnarray}
Applying the boundary conditions at $x_{5}=L$, taking into account
the mass mixing terms from Eq.~(\ref{localizedmasses}) and using the
procedure defined in Eqs.~(\ref{limit}) and (\ref{limit2}), we
derive the conditions on $f(L,h)$:
\begin{eqnarray}
\label{f.IR.bc}
\begin{array}{ccccc}
f_{1,R}^{1,...,4} + M_{B_2} f_{3,R}^{1,...,4}=0 &\quad&
f_{1,R}^{5}+M_{B1}f_{2,R}^{5}=0&\quad& f_{2,L}^{1,...,4}=0
\\
&&&&\\
 f_{3,L}^{1,...,4}-M_{B_2}f_{1,L}^{1,...,4}=0 &\quad& f_{2,L}^{5}-M_{B_1}
f_{1,L}^{5}=0&\quad& f_{3,L}^{5,...,10}=0
\end{array}
\end{eqnarray}
where the superscripts denote the vector components.

Asking that the determinant of this system of equations vanishes so
that we get a non-trivial solution, we notice that all dependence on
the prefactors $a^{-3/2}(x_5)\exp[\pm M x_{5}]$ relating $S_{\pm
M},\dot{S}_{\pm M}$ and $\tilde{S}_{\pm M},\dot{\tilde{S}}_{\pm M}$
drops out, and thus we conclude that one of the following equations
should vanish:
\begin{eqnarray}
&\tilde{S}^{'3}_{-M_{2}}=0&\label{exoticKK1}\\
&\tilde{S}_{-M_{3}}^{\prime 5}=0&\label{exoticKK2}\\
&\left[M_{B_2}^2\tilde{S}_{M_{1}}\tilde{S}_{-M_{3}}
-\frac{a^2_L}{z^2}
\tilde{S}'_{M_{1}}\tilde{S}'_{-M_{3}}\right]^2=0&\label{exoticKK3}\\
&2\tilde{S}_{M_{3}}\left[M_{B_2}^2\tilde{S}_{-M_{3}}\tilde{S}'_{-M_{1}}
+ \tilde{S}_{-M_{1}}\tilde{S}'_{-M_{3}}\right]
-M_{B_2}^2\tilde{S}'_{-M_{1}}\sin^2\left(\frac{\lambda_F
h}{f_h}\right)=0&\label{bottomKK}
\end{eqnarray}

\begin{eqnarray}
&2\left[M_{B_1}^2\tilde{S}_{M_{1}}\left(-1 +
\tilde{S}_{M_{2}}\tilde{S}_{-M_{2}}\right)
\left(M_{B_2}^2\tilde{S}_{-M_{3}}\tilde{S}'_{-M_{1}} +
      \tilde{S}_{-M_{1}}\tilde{S}'_{-M_{3}}\right)+\right.&\nonumber\\
&\left. \tilde{S}_{M_{2}}\tilde{S}'_{-M_{2}}\left(M_{B_2}^2\left(-1
+ \tilde{S}_{M_{1}}\tilde{S}_{-M_{1}}\right)\tilde{S}_{-M_{3}}-
\frac{a^2_L}{z^2}\tilde{S}_{-M_{1}}
      \tilde{S}'_{M_{1}}
      \tilde{S}'_{-M_{3}}\right)\right]+&\nonumber\\
& \left[M_{B_2}^2\tilde{S}_{M_{2}}\tilde{S}_{-M_{3}}\tilde{S}'_{-M_{2}} +
M_{B_1}^2\left(2M_{B_2}^2\tilde{S}_{M_{1}}\tilde{S}_{-M_{3}}
\tilde{S}'_{-M_{1}} +
\tilde{S}'_{-M_{3}}
+2\tilde{S}_{M_{1}}\tilde{S}_{-M_{1}}\tilde{S}'_{-M_{3}}-\right.\right.&
\nonumber\\
&\left.\left.\tilde{S}_{M_{2}}\tilde{S}_{-M_{2}}\tilde{S}'_{-M_{3}}\right)
\right]\sin^2\left(\frac{\lambda_F
h}{f_h}\right)-M_{B_1}^2\tilde{S}'_{-M_{3}}\sin^4\left(\frac{\lambda_F
h}{f_h}\right)=0&\label{topKK}
\end{eqnarray}
where for simplicity we did not write the dependence on L and z and
furthermore, we have used the Crowian:
\beq \label{e.cr}
-\dot{\tilde{S}}_M(x_5,z)\dot{\tilde{S}}_{-M}(x_5,z) +
\tilde{S}_M(x_5,z) \tilde{S}_{-M}(x_5,z) = 1 \, . \eeq
In the following, and for simplicity, we shall omit the tildes and
refer to the solutions $\tilde{S}_{\pm M}$ by $S_{\pm M}$.

The same value $\lambda_{F}=1/\sqrt{2}$ is found in the fermionic
case, so from now on $\lambda=\lambda_{G}=\lambda_{F}$. The
solutions to Eqs.~(\ref{exoticKK1}), (\ref{exoticKK2}) and
(\ref{exoticKK3}) can be interpreted as the KK spectrum of exotic
fermions. We shall name these states $E_3$, $E_2$ and $E_1$,
respectively. Eq.~(\ref{bottomKK}) is the KK mass spectrum for the
bottom quark and Eq.~(\ref{topKK}) gives the KK spectrum for the top
quark.

We can rewrite Eqs.~(\ref{bottomKK}) and (\ref{topKK}) in the same
form as the gauge bosons:

 \begin{eqnarray}
& &
 1 +  F_{b}(m_n^2)\sin^2\left(\frac{\lambda h}{f_h}\right) =0 \,, \,\label{bottomform}\\
& &
 1 +  F_{t_1}(m_n^2)\sin^2\left(\frac{\lambda h}{f_h}\right) +F_{t_2}(m_n^2)\sin^4\left(\frac{\lambda h}{f_h}\right) =0\label{topforms}
\end{eqnarray}

where
\begin{eqnarray}
F_{b}(z^2) &=&
-\frac{M_{B_2}^2S'_{-M_{1}}}{2S_{M_{3}}(M_{B_2}^2S_{-M_{3}}S'_{-M_{1}}
+ S_{-M_{1}}S'_{-M_{3}})}\\
F_{t_1}(z^2) &=&\frac{F_{1}(z^2)}{F_{d}(z^2)}\\
F_{t_2}(z^2)&=&\frac{F_{2}(z^2)}{F_{d}(z^2)},
\end{eqnarray}

\begin{eqnarray}
F_{1}(z^2)&=&M_{B_2}^2S_{M_{2}}S_{-M_{3}}S'_{-M_{2}} +
M_{B_1}^2\left[2M_{B_2}^2S_{M_{1}}S_{-M_{3}}S'_{-M_{1}} +
\right.\nonumber\\
&&\left.S'_{-M_{3}}
+2S_{M_{1}}S_{-M_{1}}S'_{-M_{3}}-S_{M_{2}}S_{-M_{2}}S'_{-M_{3}}\right]\\
F_{2 }(z^2)&=&-M_{B_1}^2S'_{-M_{3}}\\
 F_{d}(z^2)&=&
2\left[M_{B_1}^2S_{M_{1}}\left(-1 +
S_{M_{2}}S_{-M_{2}}\right)\left(M_{B_2}^2S_{-M_{3}}S'_{-M_{1}} +
      S_{-M_{1}}S'_{-M_{3}}\right)+\right.\nonumber\\
      &&\left.S_{M_{2}}S'_{-M_{2}}\left(M_{B_2}^2\left(-1 +
S_{M_{1}}S_{-M_{1}}\right)S_{-M_{3}}-
\frac{a^2_{L}}{z^2}S_{-M_{1}}
      S'_{M_{1}}
      S'_{-M_{3}}\right)\right]
\end{eqnarray}
Equations (\ref{gsf}), (\ref{bottomform}) and (\ref{topforms}) are
the starting point for computing the one-loop effective potential
for the scalars $h^\ha$.

It is physically very illuminating to interpret the origin of the
fermionic resonances if we contemplate the fermionic determinant in
the case $h=0$:
\begin{eqnarray}
\begin{array}{rrr}
S_{M_{3}}=0 &\quad& S_{M_{2}}S'_{M_{1}} +
M_{B_1}^2S_{M_{1}}S'_{M_{2}}=0\label{hzero1}\\
&&\\
S^{'5}_{-M_{3}}=0 &\quad& (M_{B_2}^2S_{-M_{3}}S'_{-M_{1}} + S_{-M_{1}}S'_{-M_{3}})^2=0\label{hzero2}\\
&&\\
S^{'4}_{-M_{2}}=0 &\quad&(M_{B_2}^2S_{M_{1}}S_{-M_{3}}
-\frac{a^2_L}{z^2}S'_{M_{1}} S'_{-M_{3}})^2=0 \label{hzero3}
\end{array}
\end{eqnarray}
We recall that
the way we defined the initial fermionic functions ($S_{\pm M}$) was
via the boundary condition on the UV brane and the chirality.
However, since we are in the $h=0$ gauge, we can use the boundary
conditions on the IR brane to further give us the final equations
the functions have to satisfy:
\begin{eqnarray}
\begin{array}{lll}
(-,-)_{L}\Rightarrow S_{M}(L)=0, && (-,-)_{R}\Rightarrow S_{-M}(L)=0 \\
(-,+)_{L}\rightarrow S_{M} \rightarrow (+,-)_{R}\Rightarrow
S'_{M}(L)=0, &\quad& (-,+)_{R}\rightarrow S_{-M} \rightarrow
(+,-)_{L}\Rightarrow S'_{-M}(L)=0
\end{array}
\end{eqnarray}
We can now interpret  Eq.~(\ref{hzero1}) in a simple way: the first
equation in the first line provides the mass spectrum for the
singlet right handed bottom. This with the first equation in the
second line provide the mass spectrum of the triplet components
$T_{1R}$, $T_{2R}$ of  $\xi_{3R}$. The second equation in the first
line may be identified with the mass spectrum of the singlet $t_{R}$
and its modification from the mixing mass term with $u'_{L}$. It can
be shown from this equation that the light top resonances will be
mostly the KK modes of the right handed top. From the first equation
in the third line we get the mass spectrum for $Q_{2R}$.
Additionally we can also see here the mixing of the bidoublets
$Q_{1}$ and $Q_{3}$. The second equation in the second line is
related to the mass spectrum for the $(t_{L},b_{L})^{t}$ doublet and
its mixing with $(q^{\prime \prime u}_{3L}, q^{\prime \prime
d}_{3L})^{t}$. The second equation in the third line may be
identified as the spectrum of a new fermionic state coming from the
doublet-doublet mixing between $(\chi^{u}_{1L},\chi^{d}_{1L})^{t}$
and $(\chi^{u}_{2L},\chi^{d}_{2L})^{t}$.

\subsection{Top Mass and Yukawa Coupling at Low Energy}

At small momenta, the solution to Eq.~(\ref{SM}) takes the form:
\begin{equation}
\tilde{S}_{M}\approx z\int^{x_{5}}_{0}
a^{-1}(x_{5})e^{-2My}dy+\mathcal{O}(z^{3}) \, .
\end{equation}
Replacing this in Eq.~(\ref{topKK}), keeping only quadratic terms in
z and for simplicity neglecting the small $M_{B{2}}$ effects, we
find that the top quark spectral equation reduces to:
\begin{equation}
\label{top.mass}
A\left(\frac{z}{\tilde{k}}\right)^{2}+M_{B_{1}}^{2}\left[1+B\left(\frac{z}{\tilde{k}}\right)^{2}\right]\sin^2\left(\frac{\lambda
h}{f_h}\right)-M_{B_{1}}^{2}\sin^4\left(\frac{\lambda
h}{f_h}\right)=0,
\end{equation}
where A and B are coefficients given by:
\begin{eqnarray}
A=-\frac{(\frac{1}{2}+c_{1})+M_{B_{1}}^{2}(\frac{1}{2}+c_{2})}{2(\frac{1}{4}-c_{1}^{2})(\frac{1}{2}+c_{2})}\quad;&\qquad&
B=\frac{(\frac{1}{4}-c_{2}^{2})-\frac{1}{2}(\frac{1}{4}-c_{1}^{2})}{2(\frac{1}{4}-c_{1}^{2})(\frac{1}{4}-c_{2}^2)}.
\end{eqnarray}
Solving Eq.~(\ref{top.mass}) for $z^2$, we find that the top mass
takes the form
\begin{equation}
\label{toplow} 
\left(\frac{m_t}{\tilde{k}}\right)^{2} \approx
-\frac{M_{B_{1}}^{2}\sin^2\left(\frac{\lambda
h}{f_h}\right)\cos^2\left(\frac{\lambda
h}{f_h}\right)}{A+M_{B_{1}}^{2}B\sin^2\left(\frac{\lambda
h}{f_h}\right)}+\co(m_t^3/\tilde{k}^3) ,
\end{equation}
which can also be written as
\begin{equation}
\left(\frac{m_t}{\tilde{k}}\right)^{2} \approx
\frac{2M_{B_{1}}^{2}\sin^2\left(\frac{\lambda
h}{f_h}\right)\cos^2\left(\frac{\lambda
h}{f_h}\right)(\frac{1}{4}-c_{1}^{2})(\frac{1}{4}-c_{2}^{2})}{(\frac{1}{2}-c_{2})(\frac{1}{2}+c_1)+M_{B_{1}}^{2}\left[(\frac{1}{4}-c^2_2)\cos^2\left(\frac{\lambda
h}{f_h}\right)+\frac{1}{2}(\frac{1}{4}-c_1^2)\sin^2\left(\frac{\lambda
h}{f_h}\right)\right]}.
\end{equation}

This expression is similar to the one obtained in a related model in
Ref.~\cite{Contino:2006qr}. From this expression we can calculate
the Yukawa coupling of the top, $Y_{t}$, at linear order, by simply
taking the derivative of $m_t$ with respect to $h$:
\begin{equation}
\label{HTT} Y_{t}=\sqrt{2}\;\frac{\partial m_{t} }{\partial
h}\simeq\frac{m_{t}g}{m_{W}\sqrt{2}}\frac{\cos\left(\frac{2\lambda
h}{f_h}\right)}{\cos\left(\frac{\lambda h}{f_h}\right)},
\end{equation}
where we note that in the limit $\lambda h/f_{h}\ll 1$, we recover
the SM top Yukawa coupling.

\section{Coleman--Weinberg potential for $h^{\ha}$ in 5D}

The Coleman--Weinberg potential in 5D KK theories (see for instance
\cite{Falkowski:2006vi}) takes the form:
\begin{equation}
V(h) =
\sum_{r}\pm\frac{N_{r}}{(4\pi)^{2}}\int_{0}^{\infty}dpp^{3}\log[\rho(-p^{2})].
\end{equation}
Here $N$ is the number of degrees of freedom of a given particle,
$\pm$ depends on whether we are considering bosons or fermions and
$\rho(z^{2})$ is the spectral function encoding the spectrum in the
presence of $h$. We showed in the previous section that in the 5D
framework the spectral functions encoding the KK spectrum in the
presence of the Higgs vev are given by:

\begin{eqnarray}
\begin{array}{lcl}
 \rho_{W}(z^2) = 1 +  F_{W}(z^2) \sin^2\left(\frac{\lambda h}{f_h}\right)&\quad& \rho_{Z}(z^2) = 1 +  F_{Z}(z^2)\sin^2\left(\frac{\lambda h}{f_h}\right) ,\\
&&\\
 \rho_{b}(z^2) = 1 +  F_{b}(z^2) \sin^2\left(\frac{\lambda
h}{f_h}\right)&\quad&\rho_{t}(z^2) = 1 +
F_{t_1}(z^2)\sin^2\left(\frac{\lambda h}{f_h}\right)
+F_{t_2}(z^2)\sin^4\left(\frac{\lambda h}{f_h}\right)
\end{array}
\end{eqnarray}

Thus, the scalar potential for the pseudo-Goldstone can be written
as:
\begin{eqnarray} \label{e.hcw}
 V(h) &=&\int^{\infty}_{0}dpp^{3}\left.\Big(-
\frac{12}{(4\pi)^2}\left.\Big\{\log\Big[1+F_{t_1}(-p^2)\sin^2\left(\frac{\lambda h}{f_h}\right) +F_{t_2}(-p^2)\sin^4\left(\frac{\lambda h}{f_h}\right)\Big]+\right.\right.\nonumber\\
& & \left.\left. \log\Big[1+ F_{b}(-p^2) \sin^2\left(\frac{\lambda
h}{f_h}\right)\Big]\right.\Big\}
+\frac{6}{(4\pi)^2}\log\Big[1+ F_{W}(-p^2) \sin^2\left(\frac{\lambda h}{f_h}\right)\Big]+\right.\nonumber\\
& & \left.\frac{3}{(4\pi)^2}\log\Big[1+ F_{Z}(-p^2)
\sin^2\left(\frac{\lambda h}{f_h}\right)\Big]\right.\Big)
\end{eqnarray}

The form factors $F_{W}$ and $F_{Z}$ for the gauge bosons are given
by Eq.~(\ref{gsf}), and $F_{b}$, $F_{t_1}$ and $F_{t_2}$ for the
fermions are given by Eq.~(\ref{bottomform}) and (\ref{topforms}).
We note here that the main contribution to the potential from the
fermionic sector is due to the top quark. Therefore, the first two
fermion multiplets, $\xi_1$ and $\xi_2$ are the most important for
fixing the Higgs potential.
\section{Numerical Results}
\label{s.d}

%
\begin{figure}[!t]\centering
\scalebox{0.9}[0.8]{\includegraphics[bb=2.9cm 0cm 18.1cm 16.1cm,
clip=false]{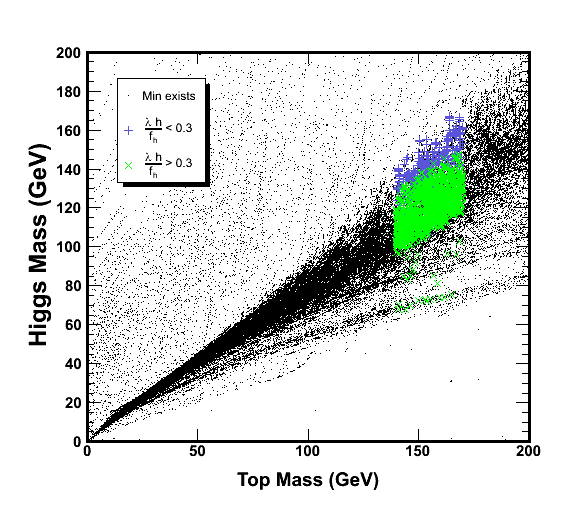}}\caption{\footnotesize{Higgs Mass vs top
mass in GeV. Blue (dark gray) crosses represent the linear regime,
green (light gray) x's the non-linear regime and black dots where a
minimum for the effective potential exists.}}
\label{figure:tophiggs}
\end{figure}
\begin{figure}[!tb]\centering
\scalebox{0.9}[0.8]{\includegraphics[bb=2.9cm 0cm 18.1cm 16.1cm,
clip=false]{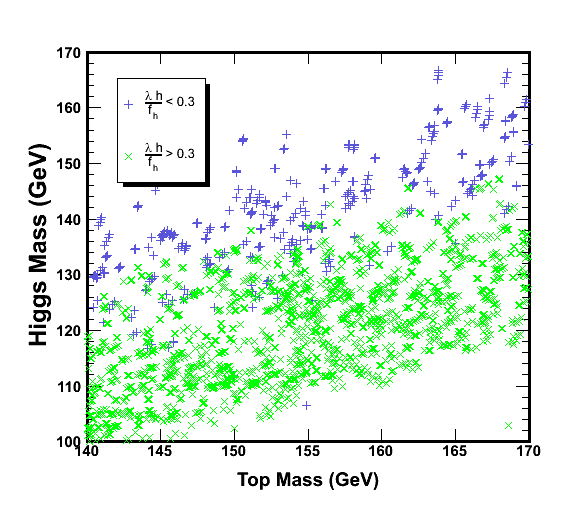}}\caption{\footnotesize{Higgs Mass vs
top mass in GeV, zoomed in region. Blue (dark gray) crosses
represent the linear regime, green (light gray) x's the non-linear
regime.}} \label{figure:higgstop}
\end{figure}
\begin{figure}[!tb]\centering
\scalebox{0.9}[0.8]{\includegraphics[bb=2.9cm 0cm 18.1cm 16.1cm,
clip=false]{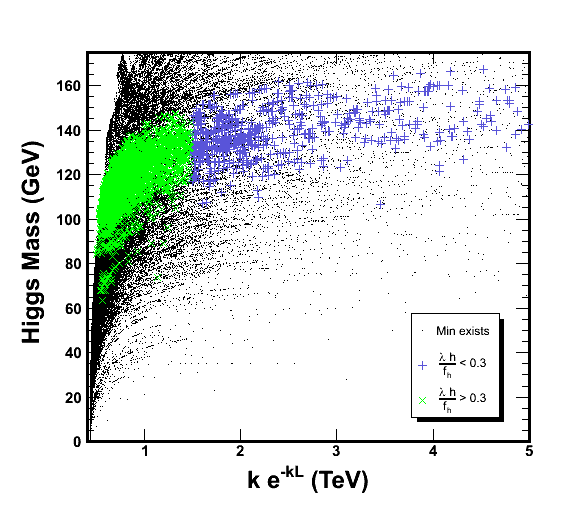}}\caption{\footnotesize{Higgs Mass (GeV) vs
$\tilde{k}$ (TeV). Blue (dark gray) crosses represent the linear
regime, green (light gray) x's the non-linear regime and black dots
where a minimum for the effective potential
exists.}}\label{figure:kaHiggs}
\end{figure}
\begin{figure}[!tb]\centering
\scalebox{0.9}[0.8]{\includegraphics[bb=2.9cm 0cm 18.1cm 16.1cm,
clip=false]{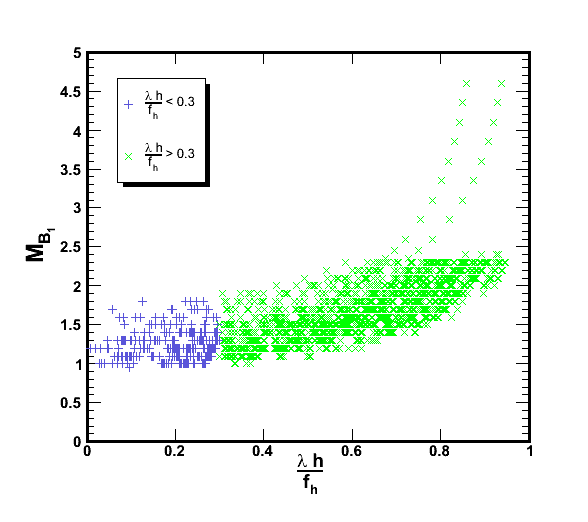}}\caption{\footnotesize{Minimum vs $M_{B_1}$.
Blue (dark gray) crosses represent the linear regime, green (light
gray) x's the non-linear regime. The sparse region for higher values
of $M_{B_1}$ is due to a coarser grid scanned in that
region.}}\label{figure:hB1}
\end{figure}
\begin{figure}[!tb]\centering
\scalebox{0.9}[0.8]{\includegraphics[bb=2.9cm 0cm 18.1cm 16.1cm,
clip=false]{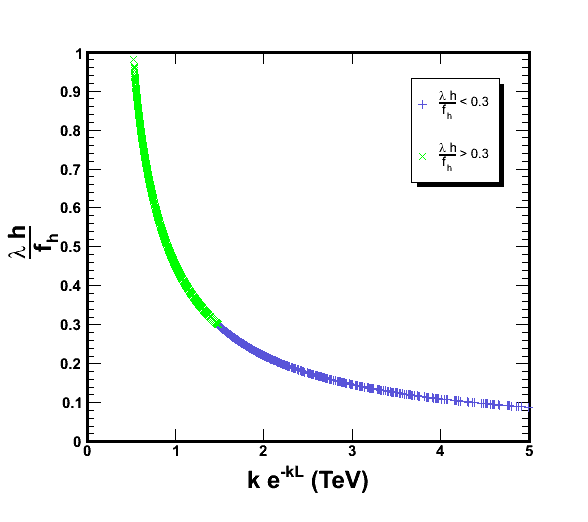}}\caption{\footnotesize{Minimum vs $\tilde{k}$
(TeV). Blue (dark gray) crosses represent the linear regime, green
(light gray) x's the non-linear regime}}\label{figure:kah}
\end{figure}
\begin{figure}[!tb]\centering
\scalebox{0.9}[0.8]{\includegraphics[bb=2.9cm 0cm 18.1cm 16.1cm,
clip=false]{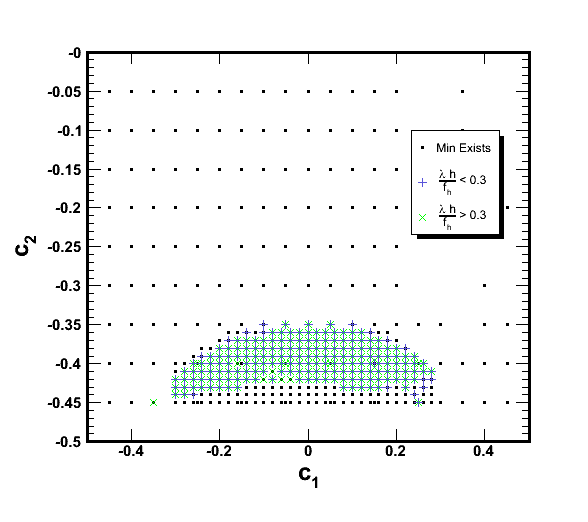}}\caption{\footnotesize{$c_1$ vs $c2$. Blue
(dark gray) crosses represent the linear regime, green (light gray)
x's the non-linear regime and black dots where a minimum for the
effective potential exists.}}\label{figure:c1c2}
\end{figure}
\begin{figure}[!tb]\centering
\scalebox{0.9}[0.8]{\includegraphics[bb=2.9cm 0cm 18.1cm 16.1cm,
clip=false]{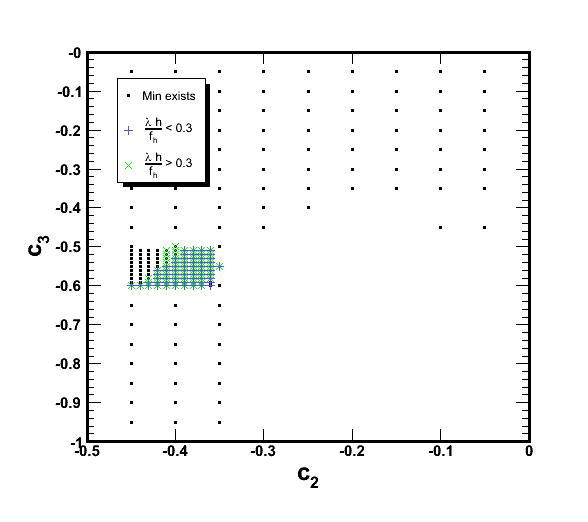}} \caption{\footnotesize{$c_2$ vs. $c_3$. Blue
(dark gray) crosses represent the linear regime, green (light gray)
x's the non-linear regime and black dots where a minimum for the
effective potential exists.}}\label{figure:c2c3}
\end{figure}
\begin{figure}[!tb]\centering
\scalebox{0.9}[0.8]{\includegraphics[bb=2.9cm 0cm 18.1cm 16.1cm,
clip=false]{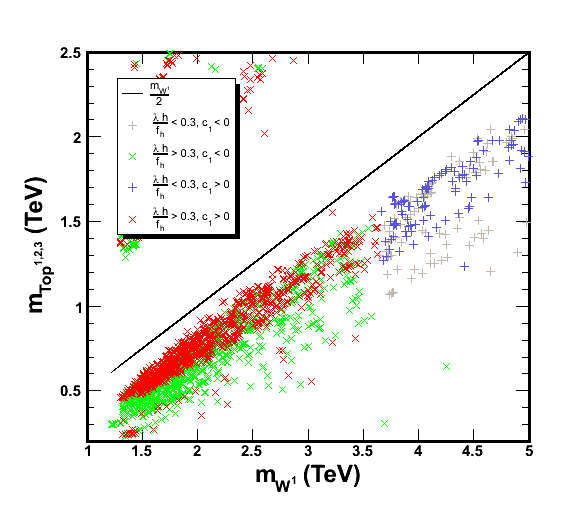}}\caption{\footnotesize{$m_{W^1}$ vs
$m_{Top^{1,2,3}}$ in TeV. Also marked is $m_{W^1}/2$ showing that
only the first excited top mode can decay into gauge bosons. Blue
(dark gray) crosses represent the linear regime with $c_{1}>0$, gray
 (light gray) crosses the linear regime with $c_{1}<0$, red x's (dark gray) the
non-linear regime with $c_{1}>0$, green x's (light gray) the
non-linear with $c_{1}<0$.}}\label{figure:W1T1}
\end{figure}

\begin{figure}[!tb]
\centering \scalebox{0.9}[0.8]{\includegraphics[bb=2.9cm 0cm 18.1cm
16.1cm, clip=false]{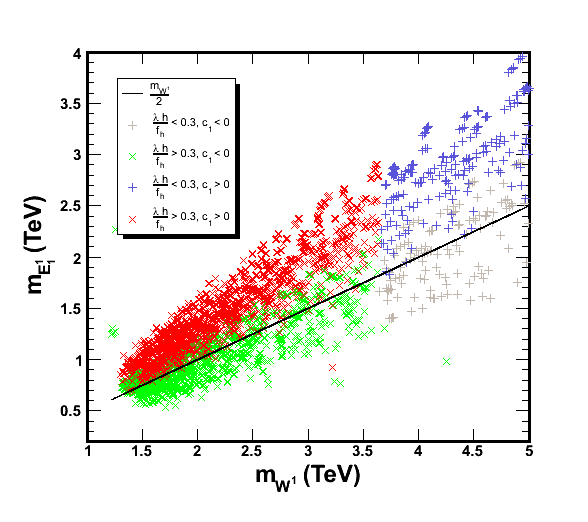}} \caption{\footnotesize{$m_{W^1}$ vs
$m_{E_1^{1,2}}$ in TeV. Also marked is $m_{W^1}/2$ showing that
depending on the value of the parameters ($c_i$ and $B_i$) the first
mode of the lightest exotic fermion may decay into the gauge bosons.
Blue (dark gray) crosses represent the linear regime with $c_{1}>0$,
gray (light gray) crosses the linear regime with $c_{1}<0$, red x's
(dark gray) the non-linear regime with $c_{1}>0$, green x's (light
gray) the non-linear with $c_{1}<0$.}\label{figure:W1E1}}
\end{figure}

Our goal is to find a non-trivial  minimum of the Higgs effective
potential that will break $SU(2)_{L}\times U(1)_{Y}$ down to
$U(1)_{EM}$. The relevant parameters which control the model are
$M_{B_1}$, $M_{B_2}$, $c_{1}$, $c_{2}$, $c_{3}$ and $k L$, where
$M_{1}=-c_{1}k$, $M_{2}=-c_{2}k$, $M_{3}=-c_{3}k$ and $L$ is the
length of the extra dimension. Looking at the potential we
immediately see that $\sin( \lambda h/f_h)=1$ is a non-trivial
extremum. When $(\lambda h/f_h)=0$ corresponds to a maximum, this
extremum can be a minimum. However this would lead to a non-standard
effective low-energy theory, since as seen from the form factors,
and from Eq.~(\ref{HWW}), the Higgs coupling to the gauge bosons
vanishes at linear order\footnote{Couplings of higher order in Higgs
field do not vanish, but the interpretation of such a theory goes
beyond the scope of this work}. Therefore, we shall concentrate only
in the regions of parameter space such that $0<\sin (\lambda
h/f_h)<1$ with $0<\lambda h/f_h<\pi/2$. After obtaining a
non-trivial minimum, we calculate $k$ by finding the first zero of
Eq.~(\ref{gsf}) and setting this mass equal to the $W$
mass~\cite{Yao:2006px}. This is the mass that the $W$ zero-mode gets
from interacting with the Higgs vev. Since the relationship between
the $Z$ and the $W$ form factors remains the same as in the SM, we
also recover a consistent mass for the $Z$ boson.

Having obtained $k$, we calculate the top-quark mass, given by the
first zero of Eq.~(\ref{topforms}), and demand that its value is in
the phenomenological range $m_{top}\in [140,170]$ GeV. We take such
a wide range since we consider the result of our calculation to be
associated with the running top-quark mass evaluated at a scale of
the order of the weak scale. Due to the strong gauge coupling, the
running top-quark mass varies rapidly between values of $m_{top}
\sim 140$~GeV, for a scale of the order of a few TeV, to values of
$m_{top} \sim 165$~GeV, for a scale of order of the top quark pole
mass $M_t$. Furthermore, we also calculate the bottom quark mass,
given by the first zero of Eq.~(\ref{bottomform}), and demand it to
be in the phenomenological range $m_{bottom}\in [2,4]$ GeV. The
Higgs mass is given by the second derivative of the effective
potential with respect to $h$, evaluated at the vev. As we will
show, somewhat smaller Higgs boson mass values tend to be associated
with smaller values of the top quark mass, a reflection of the power
dependece of the low-energy Higgs quartic coupling on the top quark
mass. The Higgs mass and the KK spectrum are seen to be only weakly
dependent on variations of the bottom quark mass in the above
defined range.

Although we will display all results consistent with the observed
quark and gauge boson low energy spectrum, we will concentrate on
points in parameter space in the linear regime, such that $\lambda
h/f_h<0.3$. By linear regime we mean the regime where higher order
interactions involving the Higgs field, $\propto(H/\tilde{k})^{n}$
with $n>1$, can be neglected and, furthermore, the couplings of the
Higgs with gauge bosons and fermions are very close to the ones in
the SM. Therefore, in this regime the direct search LEP bounds on
the Higgs mass apply, and $m_{H} \gtrsim 114$ GeV. In addition, the
low energy effective theory is well approximated by the SM, with
renormalizable couplings and a light Higgs boson. This allow us also
to compare our results with those of
Ref.~\cite{Carena:2006bn,Carena:2007ua} where they performed
calculations in the linear regime as defined above.

We can interpret $\tilde{k}$ as the associated mass scale of our
theory and of the order of the natural UV cut off of the low-energy,
SM-like effective theory. The scale $\tilde{k}$ is inversely
proportional to $\lambda h/f_h$ and therefore as we move away from
the linear regime, we are pushing the UV cut-off lower and
additionally causing the linear couplings of the Higgs to become
smaller. Therefore, the theory becomes increasingly
non-renormalizable, which leads to the need of taking into account
higher order KK-states (beyond zero-modes) in the calculation of
low-energy processes, thus making connection with SM predictions
harder. When $\tilde{k}<2.5$ TeV, KK bosonic and fermionic
resonances of exotic and SM fields have a chance to be detected in
the next generation colliders such as the LHC; hence, we will be
specifically interested in those regions.

We performed two major numerical parameter scans. In the first one
we scanned the parameter space for values $c_{i}\in [-1,1]$ with
$i=1,2,3$ and $M_{B_1},M_{B_2}\in [0,5]$. We used $kL=30$ in our
numerical calculations for the masses since the results are
invariant under variations of $kL\in[30,34]$~\footnote{k, obtained
from demanding a W-mass in accordance with experiments, adjusts
itself so that $\tilde{k}$ remains approximately constant}. The
first scan was done with a coarse grid and before we discuss details
of our numerical results, we would like to point out that the
effective potential is a completely well behaved function of all the
parameters, therefore, we expect that the gaps in our scanned space
are smoothly filled. We found that even though for some value of the
other parameters, non-trivial minima existed for nearly all the
regions of $c_i$ and $M_{Bi}$, we only found phenomenologically
consistent gauge bosons, top and bottom quark masses in the
following regions of parameter space: $0\leq |c_1|\leq 0.3$,
$0.35\leq |c_2| \leq 0.45$, $0.55\leq |c_3|\leq 0.6$, $1\lesssim
M_{B_1}$ and $M_{B_2}< M_{B_1}$. Though the results show a skew
symmetry between positive and negative values, we decided to
concentrate on negative values of $c_2$ and $c_3$, since
interestingly enough, this is the region which is consistent with
electroweak precision measurements~\cite{Carena:2006bn}.
Furthermore, positive values of $c_1$ lead to a smaller overlap of
the left-handed top and bottom zero modes with physics at the IR
brane, leading to a stronger suppression of potentially dangerous
flavor changing operators~\cite{Gherghetta:2000qt}. Therefore, it is
in this region where we performed a more thorough scan taking
smaller steps for the $c_{i}$.

In Figure~\ref{figure:tophiggs} we notice a clear relationship
between the Higgs and the top-quark masses, where on the plot we
have included the bigger coarse scan of parameter space. Focusing on
the interesting phenomenological region, we see in Figure
\ref{figure:higgstop} that in the linear regime we get Higgs masses
that tend to be above 115~GeV, which is above the experimental bound
from LEP, and below 160 GeV. Furthermore, there is a somewhat weak
dependence of the Higgs mass values on the precise value of the
running top quark mass. Higgs mass values closer to the current
experimental bound are obtained for the smallest values of the top
quark mass in the phenomenologically allowed range. In Figure
\ref{figure:kaHiggs}, we manifestly see that in the linear regime we
may obtain masses for the Higgs which are compatible with $1.5\;
\rm{TeV}\lesssim\tilde{k}\lesssim2.5\; \rm{TeV}$.

Figure~\ref{figure:hB1} provides us with information about the 5D
parameter $M_{B_1}$ which as increased moves the minimum to higher
values of $\lambda h/f_h$, getting into the non-linear regime. We
conclude that to remain in the linear regime, we cannot have an
arbitrarily high $M_{B_1}$ boundary mass parameter.

The behavior $f_{h}\propto 1/\tilde{k}$ is seen in
Figure~\ref{figure:kah} where, moreover, we notice that for the
linear regime, values of $\tilde{k}\gtrsim 1.5$ TeV are obtained.
This low value of $\tilde{k}$ will lead to interesting phenomenology
for the decays of light KK-fermions and KK-bosons.

In Figure~\ref{figure:c1c2} we see that a selected region of the
plane $c_{1}-c_{2}$ is suitable for minimizing the effective
potential with the experimentally observed $W$, $Z$, top and bottom
masses, and that the linear regime basically applies through out
this region. As emphasized before, $c_1 \gtrsim 0$ and $c_2\lesssim
-0.4$ are the ones preferred by consistency with electroweak
precision measurements~\cite{Carena:2006bn}. In
Figure~\ref{figure:c2c3} we noticed that the same phenomenologically
accepted region for $c_2$ in the $c_2-c_3$ plane with $-0.6\leq
c_3\leq-0.55$ is chosen and it is also homogenously covered in the
linear regime.

Figures~\ref{figure:W1T1} and \ref{figure:W1E1} are very interesting
for collider phenomenology and therefore deserve special attention.
Even though we haven't included the plot, we calculated the mass of
the $W$ first excited KK state numerically and verified the relation
$m_{W1}\approx 2.5 \; \tilde{k}$. If the masses of these fermions
are below the black line denoting $m_{W^1}/2$, the decay of
$\rm{W}^{1}$ and $\rm{Z}^{1}$ to pairs of these fermions is
kinematically available. Note here however, that for $c_1>0$,
$m_{E_1^1}$ is always greater than $m_{W^1}/2$. Moreover, these
exotic fermions are unstable. The decay products of $\rm{T}^{1}$ are
analyzed in Ref.~\cite{Carena:2007ua}. The first top KK excited
state, will mainly decay to Zt, Ht, Wb. The charged 5/3 fermions of
$E_1^1$ will necessarily decay to Wt from charge conservation,
leading to interesting phenomenology~\cite{Dennis:2007tv}. In
addition, the lightest KK quark state $T^1$, the top quark (via its
right- and left-handed couplings) and the bottom quark (via its
left-handed coupling) will be included in the KK gluon decay
products. This is a distinctive feature of this model when compared
with the predominant decay into right-handed top quarks in the
models considered in
Refs.~\cite{Lillie:2007yh},\cite{Agashe:2006hk}.

\section{Conclusions}

In this article, we computed the one-loop Coleman Weinberg potential
for the Higgs field in a specific model of Gauge-Higgs unification
in warped extra dimensions. We chose the specific group $SO(5)
\times U(1)_X$, which allows the introduction of custodial
symmetries protecting the precision electroweak observables as well
as a Higgs field with the proper quantum numbers under the
electroweak gauge groups. As a first step, we computed the spectral
functions of the fermions and gauge bosons of the theory when the
Higgs field acquires a vev. These are then used to calculate the
effective potential for the Higgs. We demand non-trivial minima that
lead to the proper values of the gauge boson and fermion masses in
the low energy theory. This requirement leads to a selection of a
restricted region of parameters. Interestingly enough, the selected
regions of parameters coincide with the ones previously selected in
order to obtain good agreement with the precision electroweak
observables.

Our main result is the computation of the Higgs and KK mass spectra.
Demanding that the KK gauge bosons be accessible at the LHC and also
that the effective potential minimum be in the linear regime
associated with a SM-like low energy effective theory, we obtain
Higgs boson masses which are between the present experimental bound
on this quantity and about 160 GeV. This range of masses will be
tested first at the Tevatron and then at the LHC collider in the
near future.

The KK fermion spectrum also shows interesting features. We find
that there are KK fermions which become much lighter than the KK
gauge bosons. In particular, the lightest KK mode of the top
becomes light enough so that the KK gauge bosons will decay into
it. There is an additional light KK fermion, which may become
light enough for the gauge KK bosons to decay into it, which has
exotic quantum numbers under hypercharge, leading for instance to
the presence of a light fermion with charge $5/3 e$ with
interesting phenomenological properties.

\section*{Acknowledgements}
We would like to thank Debajyoti Choudhury, Ben Lillie, Arjun Menon,
Jing Shu, Tim Tait and
particularly Marcela Carena, Eduardo Ponton and Jose
Santiago for useful discussions and comments. Work at ANL is
supported in part by the US DOE, Div.\ of HEP, Contract
DE-AC02-06CH11357. This work was also supported in part by the U.S.
Department of Energy through Grant No.
DE-FG02-90ER40560.

\newpage
\appendix
{\Large{{\bf APPENDIX}}}

\section{$SO(5)$ Generators}
\label{SO5}
\renewcommand{\theequation}{A.\arabic{equation}}
\setcounter{equation}{0}  

\renewcommand{\thefigure}{A.\arabic{figure}}
\setcounter{figure}{0}  
The generators of the fundamental representation of $SO(5)$, with
$\tr[T^\alpha.T^\beta]=C(5)\delta^{\alpha\beta}$, are given by:

\begin{eqnarray}
\label{gen} T^{a_{\rm
L,R}}_{i,j}&=&-\frac{i\sqrt{C(5)}}{2}\left[\frac{1}{2}\epsilon^{abc}(\delta_i^b\delta_j^c-\delta^b_j\delta^c_i)\pm(\delta^a_i\delta^4_j-\delta^a_j\delta^4_i)\right],\nonumber\\
T^\ha_{i,j}&=&-i\sqrt{\frac{C(5)}{2}}(\delta^\ha_i\delta^5_j-\delta^\ha_j\delta^5_i),
\end{eqnarray}
where $T^\ha$ $(\ha=1,2,3,4)$ and $T^{a_{\rm L,R}}$ $(a_{\rm
L,R}=1,2,3)$ are the generators of $SO(5)/SO(4)$ and $SO(4)$
respectively.

\renewcommand{\theequation}{B.\arabic{equation}}
\setcounter{equation}{0}  

\renewcommand{\thefigure}{B.\arabic{figure}}
\setcounter{figure}{0}  

\section{Fermion Parity Assignments and Brane Masses}
\label{parity}

Since the Higgs has positive parity, in order to preserve the $Z_2$
orbifold symmetry, we need to ensure that the product of the
parities of the fermions coupled via the Higgs is always positive.
The Higgs profile is doubly exponentially suppressed on the UV
brane, therefore, we will restrict ourselves to discussion of
parities on the IR brane.

Concentrating on the parities of the first two multiplets given in
Eq.~(\ref{multiplets}), we notice that the product of the fermion
parities coupled by the Higgs is not positive. One way to fix this
would be to switch the parities for the $SO(4)$ singlet components:
\begin{equation}
\label{f.switch1} [+,-]\quad\Leftrightarrow\quad [+,+].
\end{equation}
Due to the introduction of the brane mass terms as in
Eq.~(\ref{localizedmasses}), and looking at Eq.~(\ref{f.IR.bc}) we
note that the switch in parity is exactly equivalent to taking
$M^2_{B_1}\rightarrow 1/M^2_{B_1}$. Thus the physical results
presented in this work are invariant under the switch of parity with
just the above mentioned change in $M_{B_1}$.

Another way to understand the realization of the $Z_2$ symmetry is
via the spurion formalism. We introduce a spurion field transforming
in the fundamental representation of $SO(5)$, and acquiring a very
large vev in the $SO(4)$ singlet direction, breaking the $SO(5)$
gauge symmetry  to $SO(4)$ on the IR brane. We also introduce local
fermionic degrees of freedom on the brane and write large local mass
terms involving the spurion, between these degrees of freedom and
the ones propagating in the bulk. The fermion parities are chosen to
be those preserving $Z_2$. By using a procedure similar to the one
outlined in Eqs.~(\ref{delta}),~(\ref{limit}) and (\ref{limit2}), we
can now change the bulk fermion IR boundary conditions while
preserving $Z_2$ to give the model we are considering in this work.
This mechanism would allow us to change the IR boundary conditions
not just of the $SO(4)$ singlet fields in the  $\bf 5$'s, but also
of the bidoublets in the $\bf 5$'s or in the $\bf 10$. For the case
of the $SO(4)$ singlet fields in the $\bf 5$'s, we need to add
$SO(5)$ singlet fermions on the IR brane, that couple to the $\bf
5$'s via the spurion field. For the case of the bidoublets in the
$\bf 10$ or in the  $\bf 5$'s, we need to add local $\bf 5$'s and
singlet fermions on the IR brane, and couple them appropriately to
the bulk $\bf 10$ or  $\bf 5$'s via gauge invariant masses or
couplings to the spurion field ($SO(5)$ singlet fields are necessary
to remove unwanted degrees of freedom).  In this way, one could
start with the opposite IR parities for the three bidoublets to the
ones given in Eq.~(\ref{multiplets}), thus preserving the $Z_2$
symmetry, and reach the choice of bulk fermion boundary conditions
used  in this work.

The same spurion mechanism can be used to write the brane mass terms
in Eq.~(\ref{localizedmasses}) in an $SO(5)$ invariant way. A
spurion field which is a $\bf 5$ under $SO(5)$, can be used to write
the bidoublet-bidoublet mixing brane mass term, $M_{B_2}$, between
the $\bf 5$ and the $\bf 10$ multiplets. The spurion field is
coupled to these two via a Yukawa type interaction which can be
adjusted to give $M_{B_2}\sim \mathcal{O}(1)$. This field can be the
same as the one that is used to change the parities of the fermions
as described in the last paragraph. The same procedure, but now
using a $\bf 15\equiv 14 \oplus 1$ of $SO(5)$ can be used to
generate the mass mixing terms between the singlets, $M_{B_1}$. The
$\bf 15$ required can be thought of as arising from a tensor product
of the $\bf 5$ spurion field: $\bf 5\otimes\bar{5} \equiv 10 \oplus
14 \oplus 1$. Note that this mechanism will not generate a localized
bare mass for the Higgs due to gauge invariance.

\section{Flip Parities for $Q_{2R}$ and the Introduction of $M_{B_{3}}$}
\label{flip.parity}

\renewcommand{\theequation}{C.\arabic{equation}}
\setcounter{equation}{0}  

\renewcommand{\thefigure}{C.\arabic{figure}}
\setcounter{figure}{0}  

We also studied the effects of flipping the parities of $Q_{2R}$ as
was done in the context of Ref.~\cite{Carena:2007ua}. In this case,
the parity assignments for the rest of our fermionic content remains
the same except for $Q_{2R}$ which now takes the form,
\begin{eqnarray}
\label{a.f.bc}
\begin{array}{c}
\begin{array}{ccc}
Q_{2R} &=& \begin{pmatrix}
\chi^{u_{i}}_{2R}(+,-)_{5/3} & q^{\prime {u_{i}}}_R(+,-)_{2/3} \\
\chi^{d_{i}}_{2R}(+,-)_{2/3} & q^{\prime {d_{i}}}_R(+,-)_{-1/3}
\end{pmatrix}
\end{array}
\end{array}.
\end{eqnarray}
If we keep the same boundary mass terms as in
Eq.~(\ref{localizedmasses}) and calculate the fermionic spectral
functions we obtain,
\begin{eqnarray}
&\tilde{S}^{4}_{M_{2}}=0&\\
&\tilde{S}_{-M_{3}}^{\prime 5}=0&\\
&\left[M_{B_2}^2\tilde{S}_{M_{1}}\tilde{S}_{-M_{3}} +
\dot{\tilde{S}}_{M_{1}}\dot{\tilde{S}}_{-M_{3}}\right]^2=0&\\
&2\tilde{S}_{M_{3}}\left[M_{B_2}^2\tilde{S}_{-M_{3}}\dot{\tilde{S}}_{-M_{1}}
+ \tilde{S}_{-M_{1}}\dot{\tilde{S}}_{-M_{3}}\right]
-M_{B_2}^2\dot{\tilde{S}}_{-M_{1}}\sin^2\left(\frac{\lambda_F h}{f_h}\right)=0&\\
&2\left[M_{B_2}^2\tilde{S}_{-M_{3}}\left(-\tilde{S}_{M_{2}} +
\tilde{S}_{M_{1}}\tilde{S}_{M_{2}}\tilde{S}_{-M_{1}}+M_{B_{1}}^{2}\tilde{S}_{M_{1}}\dot{\tilde{S}}_{M_{2}}\dot{\tilde{S}}_{-M_{1}}\right)+\right.&\nonumber\\
&\left.\tilde{S}_{-M_{1}}\left(\tilde{S}_{M_{2}}\dot{\tilde{S}}_{M_{1}}+M_{B_{1}}^{2}\tilde{S}_{M_{1}}\dot{\tilde{S}}_{M_{2}}\right)\dot{\tilde{S}}_{-M_{3}}\right]&\nonumber\\
&+\left(M_{B_{2}}^{2}\tilde{S}_{M_{2}}\tilde{S}_{-M_{3}}-M_{B_{1}}^{2}\dot{\tilde{S}}_{M_{2}}\dot{\tilde{S}}_{-M_{3}}\right)\sin^2\left(\frac{\lambda_F
h}{f_h}\right)=0&
\end{eqnarray}
From these expressions we expect that the only non-trivial EW
breaking vacua will be generated when $\lambda_{F} h/f_{h}=\pi/2$
which, as discussed in section 5, leads to a theory which is highly
non-renormalizable. We performed limited numerical analysis which
confirmed this expectation.

For completeness and as a reference for future work we write the
expression for the determinant in the case of a non-zero mass mixing
boundary term $M_{B_{3}}$ as in Eq.~(\ref{localizedmasses}). For
simplicity, we turn off the singlet mass mixing $M_{B_{1}}$. Keeping
parities as in our original model, we derive the following fermionic
spectral functions,
\begin{eqnarray}
&\dot{\tilde{S}}_{-M_{3}}^{ 5}=0&\\
&\left[M_{B_2}^2\tilde{S}_{M_{1}}\tilde{S}_{-M_{3}}\dot{\tilde{S}}_{-M_{2}}+M_{B_3}^2\tilde{S}_{M_{1}}\tilde{S}_{-M_{2}}\dot{\tilde{S}}_{-M_{3}}
+\dot{\tilde{S}}_{M_{1}}\dot{\tilde{S}}_{-M_{2}}\dot{\tilde{S}}_{-M_{3}}\right]=0&\\
&2\tilde{S}_{M_{3}}\left[M_{B_2}^2\tilde{S}_{-M_{3}}\dot{\tilde{S}}_{-M_{1}}\dot{\tilde{S}}_{-M_{2}}+M_{B_3}^2\tilde{S}_{-M_{2}}\dot{\tilde{S}}_{-M_{1}}\dot{\tilde{S}}_{-M_{3}}
+
\tilde{S}_{-M_{1}}\dot{\tilde{S}}_{-M_{2}}\dot{\tilde{S}}_{-M_{3}}\right]&\nonumber\\
&-M_{B_2}^2\dot{\tilde{S}}_{-M_{1}}\dot{\tilde{S}}_{-M_{2}}\sin^2\left(\frac{\lambda_F
h}{f_h}\right)=0&
\end{eqnarray}

\begin{eqnarray}
&2\tilde{S}_{M_{2}}\left[M_{B_2}^4\tilde{S}_{M_{1}}\tilde{S}_{-M_{3}}^2\dot{\tilde{S}}_{M_{1}}\dot{\tilde{S}}_{-M_{1}}\dot{\tilde{S}}_{-M_{2}}^2
+ M_{B_2}^2\tilde{S}_{-M_{3}}\dot{\tilde{S}}_{-M_{2}} \left(2
M_{B_3}^2\tilde{S}_{M_{1}}\tilde{S}_{-M_{2}}\dot{\tilde{S}}_{M_{1}}\dot{\tilde{S}}_{-M_{1}}+\right.\right.&\nonumber\\
&\left.\left.\left(-1+2\tilde{S}_{M_{1}}\tilde{S}_{-M_{1}}\right)\dot{\tilde{S}}_{M_{1}}\dot{\tilde{S}}_{-M_{2}}\right)\dot{\tilde{S}}_{-M_{3}}+\right.&\nonumber\\
&\left.
\left(M_{B_{3}}^{4}\tilde{S}_{M_{1}}\tilde{S}_{-M_{2}}^2\dot{\tilde{S}}_{M_{1}}\dot{\tilde{S}}_{-M_{1}}+M_{B_{3}}^{2}\left(-1+2\tilde{S}_{M_{1}}\tilde{S}_{-M_{1}}\right)\tilde{S}_{-M_{2}}\dot{\tilde{S}}_{M_{1}}\dot{\tilde{S}}_{-M_{2}}+
\tilde{S}_{-M_{1}}\dot{\tilde{S}}_{M_{1}}^2\dot{\tilde{S}}_{-M_{2}}^2\right)\dot{\tilde{S}}_{-M_{3}}^2\right]&\nonumber\\
&+\left(M_{B_{2}}^{4}\tilde{S}_{M_{1}}\tilde{S}_{M_{2}}\tilde{S}_{-M_{3}}^2\dot{\tilde{S}}_{-M_{2}}^2+M_{B_{2}}^{2}\tilde{S}_{-M_{3}}\dot{\tilde{S}}_{-M_{2}}\left(
\tilde{S}_{M_{2}}\dot{\tilde{S}}_{M_{1}}\dot{\tilde{S}}_{-M_{2}}\right.\right.&\nonumber\\
&\left.\left.-2M_{B_{3}}^2\tilde{S}_{M_{1}}\left(\tilde{S}_{M_{2}}(-2+\tilde{S}_{M_{1}}\tilde{S}_{-M_{1}})
\tilde{S}_{-M_{2}}-\dot{\tilde{S}}_{-M_{1}}\dot{\tilde{S}}_{M_{1}}\dot{\tilde{S}}_{M_{2}}\dot{\tilde{S}}_{-M_{2}}\right)\right)\dot{\tilde{S}}_{-M_{3}}\right.&
\nonumber\\
&\left.+
M_{B_{3}}^{2}\left(M_{B_{3}}^2\tilde{S}_{M_{1}}\tilde{S}_{-M_{2}}\left(2-2\tilde{S}_{M_{1}}\tilde{S}_{-M_{1}}+\tilde{S}_{M_{2}}\tilde{S}_{-M_{2}}\right)+
\dot{\tilde{S}}_{M_{1}}\dot{\tilde{S}}_{-M_{2}}\left(-2\tilde{S}_{M_{2}}\tilde{S}_{-M_{2}}\dot{\tilde{S}}_{M_{1}}\dot{\tilde{S}}_{-M_{1}}+\right.\right.\right. &\nonumber\\
&\left.\left.\left.
\tilde{S}_{M_{1}}\tilde{S}_{-M_{1}}\dot{\tilde{S}}_{M_{2}}\dot{\tilde{S}}_{-M_{2}}+\dot{\tilde{S}}_{M_{1}}\dot{\tilde{S}}_{M_{2}}\dot{\tilde{S}}_{-M_{1}}
\dot{\tilde{S}}_{-M_{2}}\right)\right)\dot{\tilde{S}}_{-M_{3}}^{2}\right)\sin^2\left(\frac{\lambda_F h}{f_h}\right)&\nonumber\\
&-M_{B_{3}}^{2}\dot{\tilde{S}}_{-M_{3}}(M_{B_{2}}^{2}\tilde{S}_{M_{1}}\tilde{S}_{-M_{3}}\dot{\tilde{S}}_{-M_{2}}+(M_{B_{3}}^{2}
\tilde{S}_{M_{1}}\tilde{S}_{-M_{2}}+\dot{\tilde{S}}_{M_{1}}\dot{\tilde{S}}_{-M_{2}})\dot{\tilde{S}}_{-M_{3}})\sin^4\left(\frac{\lambda_F
h}{f_h}\right)=0&
\end{eqnarray}

Due to the term proportional to $\sin^4\left(\lambda_F
h/f_h\right)$, we expect to have EWSB in this case, with non-trivial
minima that for some region of parameter space could lie in the
linear regime as in the model studied in this work.

\end{document}